\documentclass[preprint,aps,prd,superscriptaddress,nofootinbib]{revtex4}
\usepackage{epsfig}
\usepackage{graphicx}
\usepackage{dcolumn}
\usepackage{bm}

\date{}
\begin{document}
%\documentstyle[12pt]{article}\

%\documentclass[preprint]{revtex4}
% Some other (several out of many) possibilities
%\documentclass[preprint,aps]{revtex4}
%\documentclass[preprint,aps,draft]{revtex4}

%\begin{document}
\newcommand{\ds}{\displaystyle}
\newcommand{\mc}{\multicolumn}
\newcommand{\bce}{\begin{center}}
\newcommand{\ece}{\end{center}}
\newcommand{\beq}{\begin{equation}}
\newcommand{\eeq}{\end{equation}}
\newcommand{\bea}{\begin{eqnarray}}

\newcommand{\eea}{\end{eqnarray}}
\newcommand{\cont}{\nonumber\eea\bea}
\newcommand{\cl}[1]{\begin{center} {#1} \end{center}}
\newcommand{\ba}{\begin{array}}
\newcommand{\ea}{\end{array}}
%\newcommand{\arr}{\bea}

% -------------- MATH def -------------------------------
\newcommand{\ab}{{\alpha\beta}}
\newcommand{\cd}{{\gamma\delta}}
\newcommand{\dc}{{\delta\gamma}}
\newcommand{\ac}{{\alpha\gamma}}
\newcommand{\bd}{{\beta\delta}}
\newcommand{\abc}{{\alpha\beta\gamma}}
\newcommand{\eps}{{\epsilon}}
\newcommand{\lam}{{\lambda}}
\newcommand{\mn}{{\mu\nu}}
\newcommand{\mpnp}{{\mu'\nu'}}
\newcommand{\Amuu}{{A_{\mu}}}
\newcommand{\Amuo}{{A^{\mu}}}
\newcommand{\Vmuu}{{V_{\mu}}}
\newcommand{\Vmuo}{{V^{\mu}}}
\newcommand{\Anuu}{{A_{\nu}}}
\newcommand{\Anuo}{{A^{\nu}}}
\newcommand{\Vnuu}{{V_{\nu}}}
\newcommand{\Vnuo}{{V^{\nu}}}
\newcommand{\Fmnu}{{F_{\mu\nu}}}
\newcommand{\Fmno}{{F^{\mu\nu}}}

\newcommand{\abcd}{{\alpha\beta\gamma\delta}}

% Boldmath definitions

\newcommand{\bsigma}{\mbox{\boldmath $\sigma$}}
\newcommand{\btau}{\mbox{\boldmath $\tau$}}
\newcommand{\brho}{\mbox{\boldmath $\rho$}}
\newcommand{\bpipi}{\mbox{\boldmath $\pi\pi$}}
\newcommand{\bss}{\bsigma\!\cdot\!\bsigma}
\newcommand{\btt}{\btau\!\cdot\!\btau}
\newcommand{\bnabla}{\mbox{\boldmath $\nabla$}}
\newcommand{\bphi}{\mbox{\boldmath $\tau$}}
\newcommand{\bvarphi}{\mbox{\boldmath $\rho$}}
\newcommand{\bDelta}{\mbox{\boldmath $\Delta$}}
\newcommand{\bpsi}{\mbox{\boldmath $\psi$}}
\newcommand{\bPsi}{\mbox{\boldmath $\Psi$}}
\newcommand{\bPhi}{\mbox{\boldmath $\Phi$}}
\newcommand{\bnab}{\mbox{\boldmath $\nabla$}}
\newcommand{\bpi}{\mbox{\boldmath $\pi$}}
\newcommand{\btheta}{\mbox{\boldmath $\theta$}}
\newcommand{\bkappa}{\mbox{\boldmath $\kappa$}}

\newcommand{\bA}{{\bf A}}
\newcommand{\bB}{\mbox{\boldmath $B$}}
\newcommand{\bp}{\mbox{\boldmath $p$}}
\newcommand{\bk}{\mbox{\boldmath $k$}}
\newcommand{\bq}{\mbox{\boldmath $q$}}
\newcommand{\bfe}{{\bf e}}
\newcommand{\bb}{\mbox{\boldmath $b$}}
\newcommand{\br}{\mbox{\boldmath $r$}}
\newcommand{\bR}{\mbox{\boldmath $R$}}

\newcommand{\bs}{{\bf s}}
\newcommand{\bT}{{\bf T}}
\newcommand{\fph}{${\cal F}$}
\newcommand{\aph}{${\cal A}$}
\newcommand{\dph}{${\cal D}$}
\newcommand{\fpi}{f_\pi}
\newcommand{\mpi}{m_\pi}
\newcommand{\Tr}{{\mbox{\rm Tr}}}
\def\Qb{\overline{Q}}
\newcommand{\delu}{\partial_{\mu}}
\newcommand{\delo}{\partial^{\mu}}
%\newcommand{\half}{{1\over 2}}
%\newcommand{\quart}{{1\over 4}}
%
%
% ------------------ arrow mod ---------------------
\newcommand{\up}{\!\uparrow}
\newcommand{\upup}{\uparrow\uparrow}
\newcommand{\updo}{\uparrow\downarrow}
\newcommand{\uu}{$\uparrow\uparrow$}
\newcommand{\ud}{$\uparrow\downarrow$}
\newcommand{\auu}{$a^{\uparrow\uparrow}$}
\newcommand{\aud}{$a^{\uparrow\downarrow}$}
\newcommand{\pu}{p\!\uparrow}

% ------------------------------------------------------
\newcommand{\qp}{quasiparticle}
\newcommand{\sa}{scattering amplitude}
\newcommand{\ph}{particle-hole}
\newcommand{\qcd}{{\it QCD}}
\newcommand{\integ}{\int\!d}
\newcommand{\ie}{{\sl i.e.~}}
\newcommand{\etal}{{\sl et al.~}}
\newcommand{\etc}{{\sl etc.~}}
\newcommand{\rhs}{{\sl rhs~}}
\newcommand{\lhs}{{\sl lhs~}}
\newcommand{\eg}{{\sl e.g.~}}
\newcommand{\ef}{\epsilon_F}
\newcommand{\sigt}{d^2\sigma/d\Omega dE}
\newcommand{\sige}{{d^2\sigma\over d\Omega dE}}
% ----------------------- ------------------------------
\newcommand{\rpaeq}{\beq
\left ( \begin{array}{cc}
A&B\\
-B^*&-A^*\end{array}\right )
\left ( \begin{array}{c}
X^{(\kappa})\\Y^{(\kappa)}\end{array}\right )=E_\kappa
\left ( \begin{array}{c}
X^{(\kappa})\\Y^{(\kappa)}\end{array}\right )
\eeq}
\newcommand{\ket}[1]{| {#1} \rangle}
\newcommand{\bra}[1]{\langle {#1} |}
\newcommand{\ave}[1]{\langle {#1} \rangle}
\newcommand{\half}{{1\over 2}}

%\newcounter{f1}
%\newcounter{f2}
%\renewcommand{\theequation}{\thesubsection.\arabic{equation}}
%\renewcommand{\thetable}{\thesection.\arabic{table}}
\newcommand{\singlespace}{
    \renewcommand{\baselinestretch}{1}\large\normalsize}
\newcommand{\doublespace}{
    \renewcommand{\baselinestretch}{1.6}\large\normalsize}
\newcommand{\bftau}{\mbox{\boldmath $\tau$}}
\newcommand{\bfalpha}{\mbox{\boldmath $\alpha$}}
\newcommand{\bfgamma}{\mbox{\boldmath $\gamma$}}
\newcommand{\bfxi}{\mbox{\boldmath $\xi$}}
\newcommand{\bfbeta}{\mbox{\boldmath $\beta$}}
\newcommand{\bfeta}{\mbox{\boldmath $\eta$}}
\newcommand{\bfpi}{\mbox{\boldmath $\pi$}}
\newcommand{\bfphi}{\mbox{\boldmath $\phi$}}
\newcommand{\bfR}{\mbox{\boldmath ${\cal R}$}}
\newcommand{\bfL}{\mbox{\boldmath ${\cal L}$}}
\newcommand{\bfM}{\mbox{\boldmath ${\cal M}$}}
\def\dblint{\mathop{\rlap{\hbox{$\displaystyle\!\int\!\!\!\!\!\int$}}
    \hbox{$\bigcirc$}}}
\def\ut#1{$\underline{\smash{\vphantom{y}\hbox{#1}}}$}

\def\UNITY{{\bf 1\! |}}
\def\Pom{{\bf I\!P}}
\def\lsim{\mathrel{\rlap{\lower4pt\hbox{\hskip1pt$\sim$}}
    \raise1pt\hbox{$<$}}}         %less than or approx. symbol
\def\gsim{\mathrel{\rlap{\lower4pt\hbox{\hskip1pt$\sim$}}
    \raise1pt\hbox{$>$}}}         %greater than or approx. symbol
\def\beq{\begin{equation}}
\def\eeq{\end{equation}}
\def\bea{\begin{eqnarray}}
\def\eea{\end{eqnarray}}

\title{ Nonlinear $k_{\perp}$-factorization for Quark-Gluon
Dijet Production off Nuclei}

\author{N.N. Nikolaev}%
\email{N.N. Nikolaev@fz-juelich.de}
\affiliation{Institut f\"ur Kernphysik, Forschungszentrum J\"ulich, D-52425 J\"ulich, Germany}
\affiliation{L.D. Landau Institute for Theoretical Physics, 142432 Chernogolovka, Russia}
%\affiliation{L.D. Landau Institute for Theoretical Physics, Moscow 117940, Russia}
\author{W. Sch\"afer}%
\email{Wo.Schaefer@fz-juelich.de}
\affiliation{Institut f\"ur Kernphysik, Forschungszentrum J\"ulich, D-52425 J\"ulich, Germany}
\author{B.G. Zakharov}%
\email{B.Zakharov@fz-juelich.de}
\affiliation{L.D. Landau Institute for Theoretical Physics, 142432 Chernogolovka, Russia}
%\affiliation{Institut f\"ur Kernphysik, Forschungszentrum J\"ulich, D-52425 J\"ulich, Germany}
\author{V.R. Zoller}
\email{zoller@heron.itep.ru}
\affiliation{Institute of Theoretical and Experimental
Physics, 117259 Moscow, Russia}

\date{\today}%

%% \doublespace
%% {\large ~~~~~~~~~~~~~~~~~~~~~~~~~~~~~~~~~~~~~~~~~~~~~~FZJ-IKP-TH-2004-21, hep-ph/0411365}\\
        
%% \begin{center}

%% {\Large\bf Nonlinear $k_{\perp}$-factorization for Quark-Gluon
%% Dijet
%% Production off Nuclei}\\ \vspace{1cm}
%%  { \bf N.N. Nikolaev$^{a,b)}$,
%% W. Sch\"afer$^{a)}$, B.G. Zakharov$^{b)}$ and
%% V.R. Zoller$^{c)}$ \medskip\\  }

%% \vspace{1.0cm} {\sl $^{a}$IKP(Theorie), Forschungszentrum
%% J{\"u}lich, D-52425 J{\"u}lich, Germany
%% \medskip\\
%% $^{b}$L.D. Landau Institute for Theoretical Physics, Moscow
%% 117940, Russia}
%% \medskip\\
%% E-mail: N.Nikolaev$@$fz-juelich.de; Wo.Schaefer$@$fz-juelich.de\vspace{1cm} \\
%% % $\UNITY$
%% {\bf Abstract\\    }
%% \end{center}

\begin{abstract}
The breaking of conventional linear $k_{\perp}$-factorization 
for hard processes in a nuclear environment is by now well established. 
Here we report a detailed derivation of the nonlinear $k_{\perp}$-factorization 
relations for the production of quark-gluon dijets. This process
is of direct relevance to dijets in the proton
hemisphere of proton-nucleus collisions at energies of the Relativistic
Heavy Ion Collider (RHIC). The major technical problem is a consistent
description of the non-Abelian intranuclear evolution of multiparton
systems of color dipoles. Following the technique developed in 
our early work [ N.N. Nikolaev, W. Sch\"afer, B.G. Zakharov and
V.R. Zoller, J.\ Exp.\ Theor.\ Phys.\  {\bf 97} (2003) 441], we reduce
the description of the intranuclear evolution of the $qgg\bar{q}$ state 
to the system of three coupled-channel equations in the space 
of color singlet 4-parton states $\ket{3\bar{3}},\quad \ket{6\bar{6}}$ and 
$\ket{15\overline{15}}$ (and their large-$N_c$ generalizations). At 
large number of colors $N_c$, the 
eigenstate $(\ket{6\bar{6}}-\ket{15\overline{15}})/\sqrt{2}$ 
decouples from the initial state  $\ket{3\bar{3}}$. The resulting
nuclear distortions of the dijet spectrum exhibit much similarity to
those found earlier for forward dijets in Deep Inelastic Scattering (DIS). Still
there are certain distinctions regarding the contribution from 
color-triplet $qg$ final states and from coherent
diffraction excitation of dijets. To the large-$N_c$ approximation,
we identify four universality classes of nonlinear $k_{\perp}$-factorization
for hard dijet production.
\end{abstract}
\maketitle

\pagebreak

%%%%%%%%    Section  1

\section{Introduction}

According to the conventional perturbative QCD (pQCD) factorization 
theorems the hard scattering cross sections are linear functionals 
(convolutions) of the appropriate parton densities
in the projectile and target \cite{Textbook}. An implicit
assumption behind these theorems is that the
parton densities in the beam and target are low and
the relevant partial wave amplitudes are small, so that 
the unitarity constraints can be ignored. In the case of hard
processes in a nuclear environment, the properly
defined partial wave amplitudes become proportional to the nuclear
thickness and, for a sufficiently heavy nucleus,
overshoot the $s$-channel unitarity bound. The unitarization makes the
nuclear partial waves a highly nonlinear functional of
the free nucleon amplitudes. Alternatively, in the pQCD language, the
unitarity constraints bring in a new dimensional scale into 
the problem - the so-called saturation scale. Important
implication of the nonlinear
unitarity relation between the free-nucleon and nuclear 
partial waves is that the properly defined density of
gluons in a nucleus becomes a nonlinear functional of
the gluon density in a free nucleon; the first discussions
of the fusion of partons in deep inelastic scattering (DIS)
off a nucleus go back to 1975
\cite{NZfusion}.

The emergence of a new large scale and the ensuing 
nonlinearity call for a revision of 
the pQCD factorization for hard processes in a
nuclear environment.  A consistent analysis of forward hard dijet production 
in DIS off nuclei revealed a striking 
breaking of linear $k_{\perp}$-factorization \cite{Gatchina,Nonlinear} confirmed 
later on in the related analysis of single-jet spectra in hadron-nucleus
collisions \cite{WolfgangDIS04,SingleJet}. Namely, following the pQCD treatment
of diffractive dijet production \cite{NZsplit,NSSdijet}, 
one can define the collective 
nuclear unintegrated gluon density such that it satisfies
the $s$-channel unitarity constraints and such that the familiar linear 
$k_{\perp}$-factorization (see e.g. the  recent reviews 
\cite{Andersson:2002cf}) would hold for the nuclear structure function
$F_{2A}(x,Q^2)$ and forward 
single-quark spectrum in DIS off nuclei 
because of their special Abelian features.
However, the dijet spectra in DIS and single-jet spectra in hadron-nucleus
collisions prove to be 
a highly nonlinear functionals of the collective nuclear gluon density.
Furthermore, the pattern of nonlinearity for single-jet spectra
was shown to depend strongly on the relevant partonic subprocess
\cite{SingleJet}. Our conclusions on the breaking of linear 
$k_{\perp}$-factorization for hard scattering off nuclei were
recently taken over by other authors \cite{Blaizot,Kovchegov,Raju}.

In this communication we extend the analysis  
\cite{Nonlinear,PionDijet,SingleJet} of the 
excitation of heavy flavor and leading quark dijets in DIS, 
$\gamma^*g_N\to Q\bar{Q}$, where
$g_N$ stands for the gluon exchanged with the nucleon, 
to the excitation of quark-gluon dijets 
(pQCD Bremsstrahlung tagged by a scattered quark) in 
the pQCD subprocess $q^*g_N \to q g$ off free nucleons and its
generalization to heavy nuclear targets. In the latter case  
multiple gluon exchanges between the involved partons 
and a nucleus are enhanced by a large nuclear radius. 
The issues are (i) to which extent  such multiple gluon exchanges 
can be described in terms of the unintegrated collective nuclear
gluon density and (ii) whether the nuclear factorization for quark-gluon
dijets in $qA$ collisions is similar to that for the quark-antiquark
dijets in DIS, i.e, in $\gamma^*A$ collisions. To
a certain extent, our answer is in the affirmative - the nonlinear
$k_{\perp}$-factorization properties for two processes 
exhibit much similarity.
Still, the two
cases differ substantially. For instance, the production of 
coherent diffractive dijets  makes about 50\% of the total cross
section in DIS but becomes marginal in $qA$ collisions. Furthermore,
the contributions from quark-gluon dijets in different color
multiplets have a very
distinct nonlinear $k_{\perp}$-factorization properties.
Also the effects of the initial state interaction change
substantially from the color-singlet $\gamma^*$ 
in DIS to the color-triplet quark in $qA$ collisions. On the
other hand, the unifying aspect is a treatment 
of the excitation of final-state color dipoles in the higher color
multiplets - color-octet in DIS and sextet and 15-plet
in $qA$ collisions.

The starting point of our analysis is the master formula 
(\ref{eq:2.12}) for the inclusive dijet spectrum. It is derived
based on the technique developed 
in \cite{SlavaPositronium,NPZcharm,Nonlinear,SingleJet} and
allows to calculate the dijet spectrum in terms of the 
$\textsf{S}$-matrices for interaction with the target nucleon or nucleus
of the color-singlet n-parton states, $n=2,3,4$. 
Within this technique, one deals with infrared-safe
quantities despite the fact that the incident parton - the quark
$q^*$ - is carrying a net color charge. The
calculation of the two-parton and three-parton $\textsf{S}$-matrices 
is the single-channel problem with the known solution
\cite{NZ91,NZ94,NPZcharm}. The
stumbling block is the calculation of the 4-body $\textsf{S}$-matrix.
In the case of the quark-gluon dijets it describes the
non-Abelian intranuclear evolution of the color-singlet
$qg\bar{q}g$ system of dipoles. It can be reduced to a
$3\times 3$ coupled-channel problem.
In our earlier work
\cite{Nonlinear} we published a full solution of the
related two-channel problem for the $q\bar{q}q\bar{q}$ system
which emerges in the description of quark-antiquark dijets
in DIS. Here we report the solution for 
the $qg\bar{q}g$ system of dipoles which has some new
features compared to the $q\bar{q}q\bar{q}$ state in DIS
\footnote{A brief discussion of the main results has been
reported elsewhere \cite{Nonuniversality}.}.
We go to fine details of this derivation - specifically,
the diagonalization
of the coupled-channel $\textsf{S}$-matrix and to the formulation
of explicit 
nonlinear  $k_{\perp}$-factorization formulas for the
dijet spectrum - for several reasons. First, the production 
of quark-gluon dijets without the soft gluon approximation 
has not been treated before. Second, regarding the color
properties of the incident and final states, it is a process
of sufficient generality to set a basis for the description 
of other pQCD processes. In conjunction with our earlier
results, it allows to formulate four universality classes
of nonlinear $k_{\perp}$-factorization. 
Third, recently the formal representation 
for the dijet cross section
similar to our master formula has been discussed by several
groups \cite{Blaizot,Kovchegov,Raju}, but these works stopped
short of the diagonalization of their counterpart of
our 4-body $\textsf{S}$-matrix. Correspondingly, they do not contain
explicit 
nonlinear  $k_{\perp}$-factorization formulas.

 A very rich pattern of the
process-dependent nonlinear $k_{\perp}$-factorization emerges
from the studies presented here and reported in
\cite{Nonlinear,PionDijet,SingleJet,Nonuniversality}. For instance, 
it becomes increasingly clear that a heavy
nucleus cannot be described in terms of a universal collective
glue, rather the nuclear glue must be described by the
density matrix in the color space. Furthermore, the
collective glue defined for the slice of a nucleus rather
than the whole nucleus is an integral part of the 
description of excitation of color dipoles in higher
color representations. The linear
 $k_{\perp}$-factorization for the single-quark jets in DIS
found in our earlier study \cite{Nonlinear}
is an exception due to the Abelian incident parton - the photon.

From the point of view of practical applications, the
discussed quark-gluon dijets 
are of direct relevance to  the large (pseudo)rapidity 
region of proton-proton and proton-nucleus collisions at RHIC
(for the discussion of the possible upgrade of detectors at RHIC II 
for the improved coverage of the proton fragmentation region
see \cite{RHIC_II}).
Our treatment is applicable when the beam and final state
partons interact coherently over the whole longitudinal
extension of the nucleus, which
at RHIC amounts to the proton fragmentation region 
of 
\beq
x={(Q^*)^2+M_{\perp}^2 \over 2mE_{q^*}}
\lsim x_A ={1\over 2R_A m_p} \approx 0.1 A^{-1/3}\,, 
\label{eq:1.1}
\eeq
where $R_A$ is the 
radius of the target nucleus of mass number $A$, 
 $(Q^*)^2$ and $E_{q^*}$ are the virtuality and  
energy of the beam quark $q^*$ in the target rest frame,
$M_{\perp}$ is the transverse mass of the dijet
and $m_p$ is the proton mass (\cite{NZfusion,NZ91},
for the related color dipole phenomenology of the
experimental data on nuclear shadowing see \cite{BaroneShad}). 

The presentation of the main material is organized as follows.
The master formula
for the dijet spectrum is presented in Sec. II.  
The two-body density matrix - the Fourier transform of which
gives the dijet spectrum - contains the $\textsf{S}$-matrices for
the interaction of two-, three- and four-parton color-singlet
systems of dipoles with the target. Based on the technique
developed in \cite{NPZcharm}, in Sec. III we report 
single-channel $\textsf{S}$-matrices in terms of the quark-antiquark and
quark-antiquark-gluon color-dipole cross sections \cite{NZ91,NZ94}.
Sec. IV contains all the technicalities of the derivation of
the coupled-channel $\textsf{S}$-matrix for the $qg\bar{q}g$ state: the
decomposition into color multiplets; projection onto the final states;
the color-flow
diagram technique for the calculation of the $3\times 3$ matrix 
of color-dipole cross sections; the derivation of the relevant
Casimir operators; the explicit diagonalization of the 
$\textsf{S}$-matrix at large number of colors $N_c$ and the Sylvester
expansion. The quark-gluon dijet spectrum for the free-nucleon
target is derived in Sec. V. Here we also comment on a direct
relationship between the dijet and single-jet spectra 
for the free-nucleon reactions described by the 
single-gluon exchange in the $t$-channel. The principal result
of this study - the nonlinear $k_{\perp}$-factorization 
for the dijet spectrum produced off nuclear targets - is reported
in Section VI. Here we compare the pattern of nonlinear
 $k_{\perp}$-factorization for quark-gluon dijets in $qA$
collisions to that for the quark-antiquark dijets in DIS and
$gA$ collisions and identify four universality classes
depending on the color representation of the incident parton
and final-state dijet.
In Section VII we apply our results to the nuclear broadening 
of the dijet acoplanarity distribution. In Sec. VIII we comment on a 
 limiting case when the quark-gluon dijets merge to one jet. 
Such monojets can be identified with the fragmentation of the 
quark jet formed by the quasielastically scattered incident 
quark. The separation into the dijet and monojet final states 
changes with the mass number of the target nucleus and 
the centrality of the collisions. We also comment on 
the possible nuclear modification of the fragmentation function. 
In the Conclusions section we summarize our main 
results.

%%%%%%%%%%%  Section 2

\section{The master formula for quark-gluon dijet production off 
free nucleons and nuclei}

%%%%%%%%%%%%%%%%  Section 2.1

\subsection{Kinematics and nuclear coherency}
 
To the lowest order in pQCD the underlying subprocess for quark-gluon
dijet production in the proton fragmentation region of proton-nucleus
collisions is a collision of the quark $q^*$ from the proton with the 
gluon $g_N$ from the target,
$$
q^* g_N \to qg\,.
$$
It is a pQCD Bremsstrahlung tagged by a scattered quark. We don't restrict ourselves to
soft gluons.
In the case of a nuclear target one has to deal with
multiple gluon exchanges which are enhanced by a large thickness of the
target nucleus. 

\begin{figure}[!t]
\begin{center}
\includegraphics[width = 5.5cm, height= 6.0cm,angle=270]{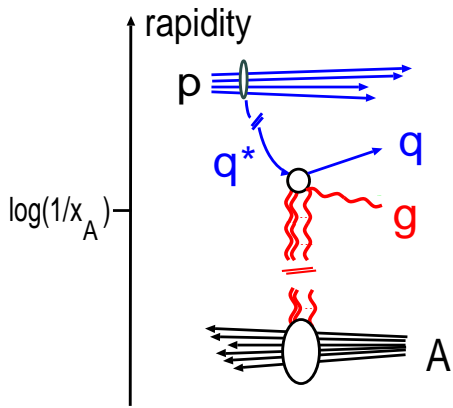}
\caption{The rapidity structure of the 
radiation of gluons by quarks $q\to qg$ in $pA$
collisions.}
\label{fig:QuarkGluonRapidity}
\end{center}
\end{figure}

From the laboratory, i.e., the nucleus rest frame, standpoint it can
be viewed as an excitation of the perturbative $|qg\rangle$ Fock state
of the physical projectile $|q^*\rangle$ by one-gluon exchange with the
target nucleon or multiple gluon exchanges with the target nucleus.
Here the collective nuclear effects develop if the coherency over
the thickness of the nucleus holds for the $qg$ Fock states, i.e.,
if the coherence length is larger than the diameter of the nucleus,
\beq
l_c ={ 2E_{q^*} \over M_{\perp}^2} = {1\over x m_N}> 2R_A\,,
\label{eq:2.1}
\eeq
where
\beq
M_{\perp}^2 = {\bp_q^2 \over z_q}+{\bp_g^2\over z_g}
\label{eq:2.2}
\eeq
is the transverse mass squared of the $qg$ state,
$\bp_{q,g}$ and $z_{q,g}$ are the transverse momenta and
fractions of the the incident quarks momentum carried by the
quark and gluon, respectively ($z_q+z_g=1$).
In the antilaboratory (Breit) frame, the partons with the momentum
$xp_N$ have the longitudinal localization of the order of their
Compton wavelength $\lambda = 1/xp_N$, where $p_N$ is the momentum
per nucleon. The coherency over the thickness of the nucleus
in the target rest frame is equivalent to the spatial overlap of
parton fields of different nucleons at the same impact parameter 
in the Lorentz-contracted ultrarelativistic nucleus. In the 
overlap regime one would think of the fusion of partons
form different nucleons and collective nuclear
parton densities \cite{NZfusion}. The overlap takes place 
if $\lambda$ exceeds the Lorentz-contracted thickness
of the ultrarelativistic nucleus,
\beq
\lambda = {1\over xp_N} > 2R_A \cdot {m_N\over p_N}\, ,
\label{eq:2.3}
\eeq
which is identical to the condition (\ref{eq:2.1}). 

Qualitatively, the both descriptions of collective nuclear effects
are equivalent to each other. Quantitatively, the
laboratory frame approach takes advantage of the 
well developed multiple-scattering theory of 
interactions of color dipoles with nuclei
\cite{NZ91,NZ92,NZ94,Nonlinear}. 
From the practical
point of view, the coherency condition $
x< x_A $ restricts collective effects in hard processes at RHIC 
to the proton 
fragmentation region. 
The target frame rapidity structure of the considered $q^*\to qg$ 
excitation 
is shown in Fig.~\ref{fig:QuarkGluonRapidity}. The (pseudo)rapidities 
of the final state partons must satisfy  $\eta_{q,g} > \eta_A=
\log{1/x_A}$. The rapidity separation of the quark and gluon hard 
jets,
\beq
\Delta\eta_{qg} = \log {1-z_g\over z_g}\,,
\label{eq:2.4}
\eeq
is considered to be finite. Both jets are supposed to be 
separated by a large rapidity from other jets at mid-rapidity
or in the target nucleus hemisphere; the gaps between all jets,
beam spectators and target debris are filled by soft hadrons
from an underlying event.

%%%%%%%%%%%%%%%  section 2.2

\subsection{Master formula for excitation of quark-gluon dijets}

In the nucleus rest frame, relativistic partons
$q^*,q$ and $g$, 
propagate along straight-line, fixed-impact-parameter, trajectories.
To the lowest
order in pQCD the Fock state expansion for
the physical state $|q^*\rangle_{phys}$ reads
\beq
 \ket{q^*}_{phys} = \ket{q^*}_0 + \Psi(z_g,\br) \ket{qg}_0\, ,
\label{eq:2.5}
\eeq
where $\Psi(z_g,\br)$ is the probability amplitude to find the $qg$ system
with the separation $\br$ in the two-dimensional impact parameter space,
the subscript $"0"$ refers to bare partons. 
The perturbative
coupling of the $q^*\to qg$ transition is reabsorbed into the lightcone
wave function $\Psi(z_g,\br)$.
We also omitted a wave function renormalization factor,
which is of no relevance for the inelastic excitation 
to the perturbative order discussed here. The explicit
expression for  $\Psi(z_g,\br)$ in terms of the quark-splitting 
function will be presented below. The wave function
depends on the virtuality of the
incident $q^*$, which equals $(Q^*)^2=(\bp^*)^2$, where $\bp^*$ is the
transverse momentum of $q^*$ in the incident proton 
(Fig.~\ref{fig:QuarkGluonRapidity}). For the
sake of simplicity we take the collision axis along the momentum of
the incident quark $q^*$, the transformation between the transverse momenta in
the $q^*$-target and $p$-target reference frames is trivial
\cite{SingleJet}.

\begin{figure}[!t]
\begin{center}
\includegraphics[width = 5.5cm,height=13cm, angle = 270]{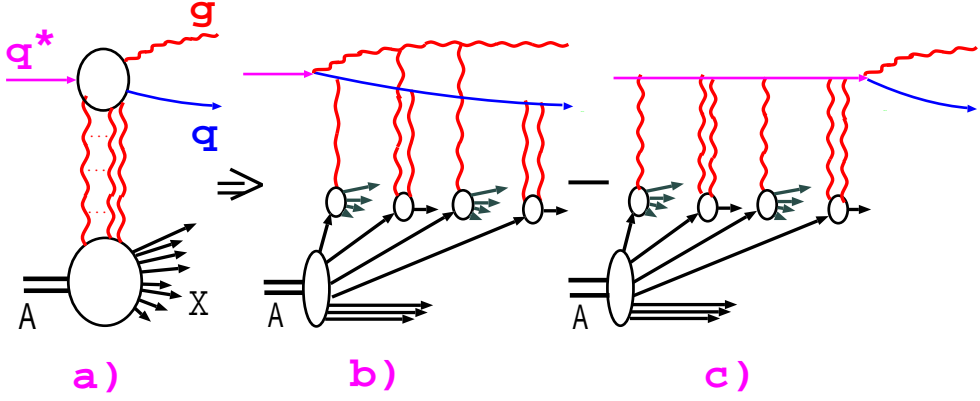}
\caption{Typical contribution to the excitation amplitude for $q^* A \to qg X$,
with multiple color excitations of the nucleus. 
The amplitude receives contributions from processes that involve interactions 
with the nucleus after and before the virtual 
decay which interfere destructively.}
\label{fig:QuarkGluonDijetExcitation}
\end{center}
\end{figure}

If $\bb$ is the impact parameter of the projectile $q^*$, then
\beq
\bb_{q}=\bb-z_g\br, \quad\quad \bb_{g}=\bb+z_q\br\, .
\label{eq:2.6}
\eeq
By the conservation of
impact parameters, the action of the $\textsf{S}$-matrix on $\ket{a}_{phys}$
takes a simple form
\bea
&& \textsf{S}\ket{q^*}_{phys} =
\textsf{S}_q(\bb) \ket{q^*}_0 +
\textsf{S}_q(\bb_q) \textsf{S}_g(\bb_g)\Psi(z,\br) \ket{qg}_0 \nonumber \\
&&=\textsf{S}_q(\bb) \ket{q^*}_{phys} +  [ \textsf{S}_q(\bb_q) \textsf{S}_g(\bb_g) - \textsf{S}_q(\bb) ]
\Psi(z_g,\br) \ket{qg} \, . 
\label{eq:2.7} \eea 
In the last line we
explicitly decomposed the final state into the (quasi)elastically
scattered $\ket{q^*}_{phys}$ and the excited state $\ket{qg}_{0}$.
The two terms in the latter describe a scattering on the target of
the $qg$ system formed way in front of the target and the
transition $q^*\to qg$ after the interaction of the state
$\ket{q^*}_0$ with the target, as illustrated in Fig. 
\ref{fig:QuarkGluonDijetExcitation}.
The contribution from transitions
$q^*\to qg$ inside the target nucleus vanishes in the high-energy
limit of $x \lsim x_A$ \footnote{In terms of the lightcone approach
to the QCD Landau-Pomeranchuk-Migdal effect, this corresponds to
the thin-target limit \cite{SlavaLPM}.}. We recall, that the $s$-channel helicity of all
partons is conserved.

\begin{figure}[!t]
\begin{center}
\includegraphics[width = 5.5cm,angle=270]{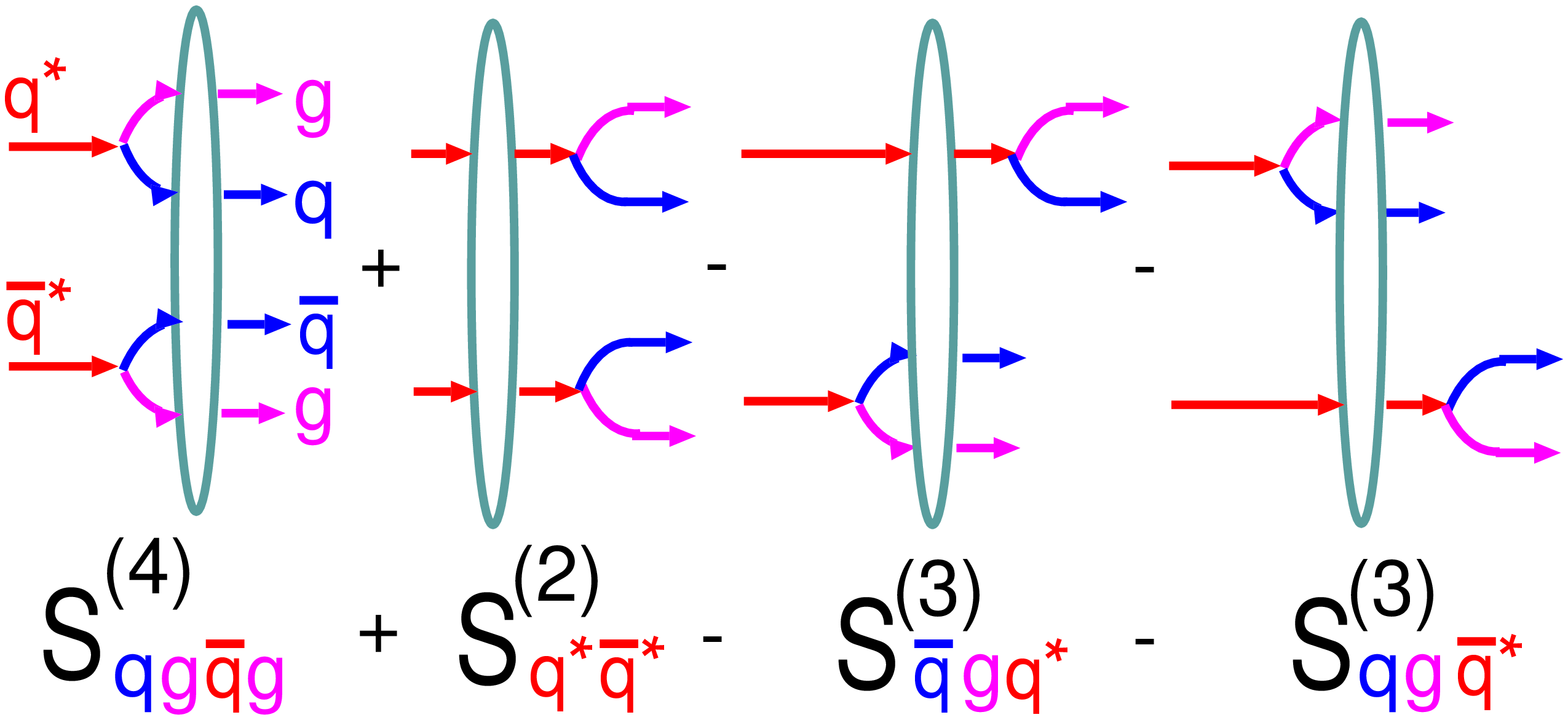}
\caption{The $\textsf{S}$-matrix structure of the two-body density
matrix for excitation $q^*\to qg$.}
\label{fig:QuarkGluonDijetDensityMatrix}
\end{center}
\end{figure}

The probability amplitude for the two-jet spectrum is given by the
Fourier transform \beq \int d^2\bb_q d^2\bb_g \exp[-i(\bp_q\bb_q +
\bp_g\bb_g)][ \textsf{S}_q(\bb_q) \textsf{S}_g(\bb_g) - \textsf{S}_q(\bb) ] \Psi(z_g,\br)
\label{eq:2.8} \eeq 
The differential cross section is proportional
to the modulus squared of (\ref{eq:2.8}),
\bea 
&&\int d^2\bb_q' d^2\bb_g' \exp[i(\bp_q\bb_q' +
\bp_g\bb_g')][ \textsf{S}_q^{\dagger}(\bb_q') \textsf{S}_g^{\dagger}(\bb_g') - \textsf{S}_q^{\dagger}(\bb') ] \Psi^{*}(z_g,\br')\nonumber\\
&\times& \int d^2\bb_q d^2\bb_g \exp[-i(\bp_q\bb_q +
\bp_g\bb_g)][ \textsf{S}_q(\bb_q) \textsf{S}_g(\bb_g) - \textsf{S}_q(\bb) ] \Psi(z_b,\br)\, .
\label{eq:2.9} \eea 
The crucial point is that 
the hermitian conjugate
$\textsf{S}^{\dagger}$ can be viewed as the $\textsf{S}$-matrix for an antiparton
\cite{SlavaPositronium,NPZcharm,Nonlinear}. Consequently,
the four terms in the product 
$$[ \textsf{S}_q(\bb_q') \textsf{S}_g(\bb_g') - \textsf{S}_{q}(\bb') ]^\dagger
[ \textsf{S}_q (\bb_q) \textsf{S}_g(\bb_g) - \textsf{S}_q(\bb) ]
$$ 
admit a simple interpretation:
\bea
\textsf{S}^{(2)}_{\bar{q^*}q^*}(\bb',\bb)&=& \textsf{S}_{q}^{\dagger}(\bb')\textsf{S}_q(\bb)
\label{eq:2.10}
\eea 
can be viewed as an $\textsf{S}$-matrix for elastic scattering on a 
target of the $\bar{q^*}q^*$ state in which the antiparton
$\bar{q}^*$ propagates at the impact parameter $\bb'$. The averaging
over the color states of the beam parton $q^*$ amounts to the
dipole 
 $q^*\bar{q^*}$ being in the color singlet state. Similarly, 
\bea
 \textsf{S}^{(3)}_{\bar{q^*}qg}(\bb',\bb_q,\bb_g) &=&
\textsf{S}_q^{\dagger}(\bb')\textsf{S}_q(\bb_q) \textsf{S}_g(\bb_g), \nonumber \\
\textsf{S}^{(3)}_{\bar{q'}g'q^*}(\bb,\bb_q',\bb_g') &=&
\textsf{S}_q^{\dagger}(\bb_q')\textsf{S}_g^{\dagger}(\bb_g') \textsf{S}_q(\bb)
 \nonumber \\
\textsf{S}^{(4)}_{\bar{q}'g' g q}(\bb_q',\bb_g',\bb_q,\bb_g) &=&
\textsf{S}_q^{\dagger}(\bb_q')\textsf{S}_g^{\dagger}(\bb_g') \textsf{S}_g(\bb_g)\textsf{S}_q(\bb_q) \,
.
\label{eq:2.11}
\eea
describe elastic scattering on a 
target of the overall
color-singlet $\bar{q}qg$ and $\bar{q}\bar{g}gq$ states,
respectively. This is shown schematically in 
Fig.~\ref{fig:QuarkGluonDijetDensityMatrix}.
Here we suppressed the matrix elements of $\textsf{S}^{(n)}$ over the
target nucleon, for details of the derivation based on the
closure relation, see \cite{Nonlinear}. Specifically, in the
calculation of the inclusive cross sections one averages over
the color states of the beam parton $q^*$, sums over color
states $X$ of final state partons $q,g$, takes the matrix products of
$\textsf{S}^{\dagger}$ and $\textsf{S}$ with respect to the relevant color indices
entering $\textsf{S}^{(n)}$ and sums over all nuclear
final states applying the closure relation.
The technicalities of the derivation
of $\textsf{S}^{(n)}$ will be presented below, here we cite
the master formula for the dijet cross section, which is the Fourier 
transform of the two-body density matrix: 
\bea
&&{d \sigma (q^* \to qg) \over dz d^2\bp_q d^2\bp_g } = {1 \over (2 \pi)^4} \int
d^2\bb_q d^2\bb_g d^2\bb'_b
 d^2\bb'_c \nonumber \\
&\times& \exp[-i \bp_q
(\bb_q -\bb'_q) - i
\bp_g(\bb_g
-\bb_g')] \Psi(z_g,\bb_q -
\bb_g) \Psi^*(z_g,\bb'_q-
\bb'_g) \\
&&\sum_{X}\bra{X}\Biggl\{
\textsf{S}^{(4)}_{\bar{q}g' qg}(\bb_q',\bb_g',\bb_q,\bb_g) 
+ \textsf{S}^{(2)}_{\bar{q^*}q^*}(\bb',\bb) -
\textsf{S}^{(3)}_{\bar{q}g'q^*}(\bb,\bb_q',\bb_g')
- \textsf{S}^{(3)}_{\bar{q^*}qg}(\bb',\bb_q,\bb_g) \Biggr\}\ket{in}\nonumber
\label{eq:2.12} 
\eea 

Hereafter, we describe the final state dijet in terms of the gluon jet 
momentum, $\bp\equiv \bp_g,~~z\equiv z_g$, 
and the decorrelation (acoplanarity) momentum
$\bDelta=\bp_q+\bp_g$. We also introduce 
\beq
\bs=\bb_q-\bb_q'\,,
\label{eq:2.13} 
\eeq
in terms of which $\bb_g-\bb_g'=\bs+\br -\br'$ and
\bea
 \exp[-i \bp_q
(\bb_q -\bb'_q) - i
\bp_g(\bb_g
-\bb_g')] =\exp[-i\bDelta \bs- i\bp \br +i\bp \br']\, ,
\label{eq:2.14} 
\eea 
so that the dipole parameter $\bs$ is conjugate to the acoplanarity 
 momentum $\bDelta$.

%%%%%%%%%%  Section 3

\section{Calculation of the 2-parton and 3-parton $\textsf{S}$-matrices}

%%%%%%%%%%%  Section 3.1
\subsection{The quark-nucleon $\textsf{S}$-matrix and the $k_{\perp}$-factorization
representations for the color dipole cross section}\

In order to set the formalism, we start with the $\textsf{S}$-matrix
representation for the cross section of interaction
of the triplet-antitriplet color dipole $q\bar{q}$ with the free-nucleon
target
\cite{Nonlinear}.
To the two-gluon exchange approximation, the $\textsf{S}$-matrices of the
quark-nucleon and antiquark-nucleon interaction equal, respectively, 
\bea 
\textsf{S}(\bb_q) & =& 1
+ iT^a V_a\chi(\bb_q)- {1\over 2}
T^aT^a \chi^2(\bb_q)\, , \nonumber\\
\textsf{S}^\dagger(\bb_{\bar{q}}) & =& 1
- iT^aV_a \chi(\bb_{\bar{q}}) - {1\over 2}T^aT^a \chi^2(\bb_{\bar{q}})\,,
\label{eq:3.1.1}
\eea
were $T^a V_a\chi(\bb)$ is the eikonal for the quark-nucleon gluon 
exchange. The vertex $V_a$ for excitation
of the nucleon $g^a N \to N^*_a$ into color octet state is so
normalized that after application of closure over
the final state excitations $N^*$ the vertex $g^a g^b
NN$ equals $\bra{ N} V_a^\dagger V_b \ket{ N}=\delta_{ab}$. 
The second order terms in (\ref{eq:3.1.1})
do already use this normalization. 
\begin{figure}[!t]
\begin{center}
\includegraphics[width = 10.5cm]{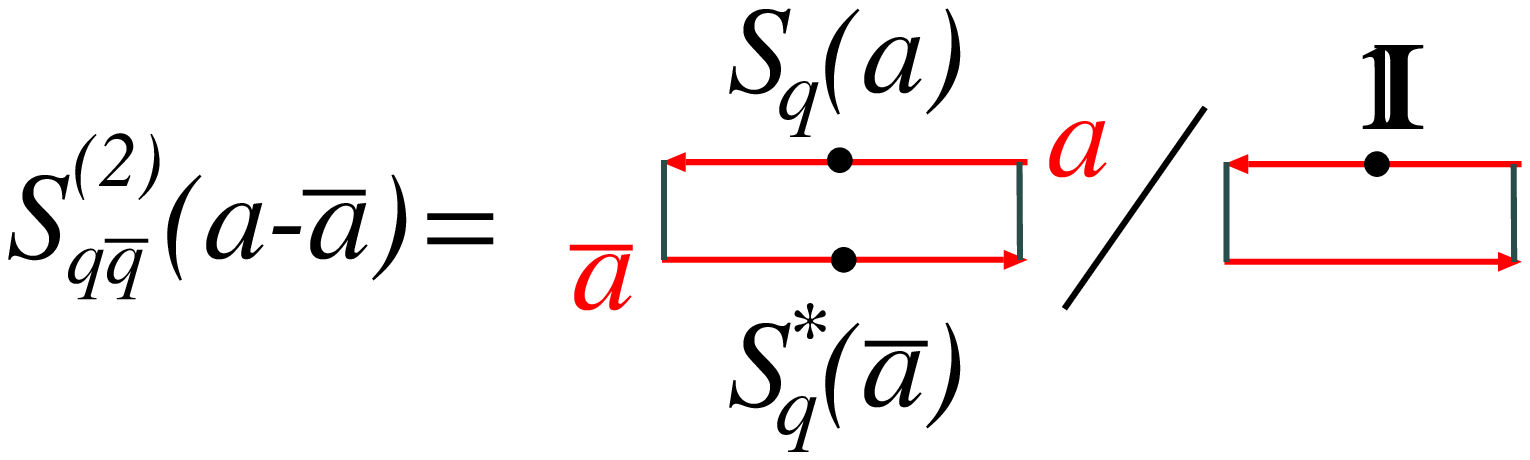}
\caption{The color-flow diagram for the $\textsf{S}$-matrix for
the interaction of the  color-single $q\bar{q}$ dipole with the
nucleon; $a$ and $\bar{a}$ are the impact parameters of
the quark and antiquark, respectively.}
\label{fig:SmatrixQbarQ}
\end{center}
\end{figure}
The $\textsf{S}$-matrix of the
$(q\bar{q})$-nucleon interaction equals
\beq
\textsf{S}^{(2)}_{q\bar{q}}(\bb_q,\bb_{\bar{q}})=
{\bra{N}{\rm Tr}[\textsf{S}(\bb_q)\textsf{S}^\dagger(\bb_{\bar{q}})] 
\ket{N} \over \bra{N}  \openone \ket{N}{\rm Tr} \openone }\,.
\label{eq:3.1.2}
\eeq
A  graphical rule for the calculation of the color traces
entering (\ref{eq:3.1.2}) is shown in Fig.~\ref{fig:SmatrixQbarQ};
such color flow diagrams will extensively be used in the 
subsequent calculation
of $\textsf{S}^{(4)}$.
 
\begin{figure}[!t]
\begin{center}
\includegraphics[width = 12.5cm]{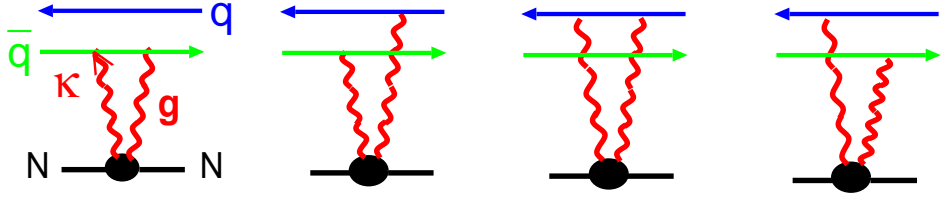}
\caption{The four Feynman diagrams for the quark-antiquark
dipole-nucleon interaction by the two-gluon pomeron exchange
in the $t$-channel.}
\label{fig:SigmaDipoleFeynman}
\end{center}
\end{figure}

The corresponding profile function is 
$\Gamma_2(\bb_q,\bb_{\bar{q}})= 1 - \textsf{S}^{(2)}_{q\bar{q}}(\bb_q,\bb_{\bar{q}})$.
The dipole cross section for interaction of the color-singlet
$q\bar{q}$ dipole $\br=\bb_q-\bb_{\bar{q}}$  with the free
nucleon
is obtained
upon the integration over the overall impact parameter 
\bea
\sigma(\br) = 2\int d^2\bb_{q} 
\Gamma_2(\bb_q,\bb_q-\br) =
C_F \int
d^2\bb_q [\chi(\bb_q)-\chi(\bb_{q}-\br)]^2\, , 
\label{eq:3.1.3} 
\eea
where $C_F=(N_c^2-1)/2N_c$ is the quark Casimir operator.
It sums a contribution from
the four Feynman diagrams of Fig. \ref{fig:SigmaDipoleFeynman}
and is related to the gluon density in the target by
the $k_{\perp}$-factorization formula \cite{NZ94,NZglue}
\bea 
\sigma(x,\br) &=& \int d^2\bkappa f (x,\bkappa)
[1-\exp(i\bkappa\br)]\,,
\label{eq:3.1.4} 
\eea
where
\bea
f (x,\bkappa)&=& {4\pi \alpha_S(r)\over N_c}\cdot {1\over \kappa^4} \cdot {\cal
F}(x,\kappa^2)
\label{eq:3.1.5} 
\eea
and
\beq
{\cal F}(x,\kappa^2) = {\partial G(x, \kappa^2) \over \partial \log \kappa^2}
 \label{eq:3.1.6}
\eeq 
is  the unintegrated gluon density in the
target nucleon. Hereafter, unless it may cause a confusion, we
suppress the variable $x$ in the gluon densities and
dipole cross sections.
The leading Log${1\over x}$ evolution of the dipole
cross section is governed by the color-dipole BFKL evolution
\cite{NZZBFKL,NZ94}, the same evolution for the
unintegrated gluon density is governed by the familiar
momentum-space BFKL equation \cite{BFKL}.

The $\textsf{S}$-matrix for coherent interaction of the color dipole
with the nuclear target is given by the Glauber-Gribov 
formula \cite{Glauber,Gribov}
\bea 
\textsf{S}[\bb,\sigma(\br)]&=&\exp[-{1\over 2}\sigma(\br) T(\bb)]\, , 
\label{eq:3.1.7} 
\eea 
where
\bea
T(\bb)= \ds \int_{-\infty}^\infty dr_z \, n_A(\bb,r_z)
\label{eq:3.1.8} 
\eea 
is the optical thickness
of the nucleus. The nuclear density $n_A(\bb,r_z)$ is
normalized according to $\ds \int d^3\vec{r} \, n_A(\bb,r_z) = \int
d^2\bb\,  T(\bb) = A$, where $A$ is the nuclear mass number.

In the specific case of $\textsf{S}^{(2)}_{\bar{q^*}q^*}(\bb',\bb)$ the 
color dipole equals
\beq
\br_{q\bar{q}}= \bb-\bb'=\bs +z\br-z\br'\, 
\label{eq:3.1.9}
\eeq 
and  $\textsf{S}^{(2)}_{\bar{q^*}q^*}(\bb',\bb)$ entering Eq.~(\ref{eq:2.12}) 
will be given by the Glauber-Gribov formula
\beq
\textsf{S}^{(2)}_{\bar{q^*}q^*}(\bb',\bb)=
\textsf{S}[\bb,\sigma(\bs +z\br-z\br')]\,.
\label{eq:3.1.10}
\eeq

%%%%%%%%%%%  Section 3.2
\subsection{The $\textsf{S}$-matrix for the color-singlet $\bar{q}qg$ state}

\begin{figure}[!t]
\begin{center}
\includegraphics[width = 10.5cm]{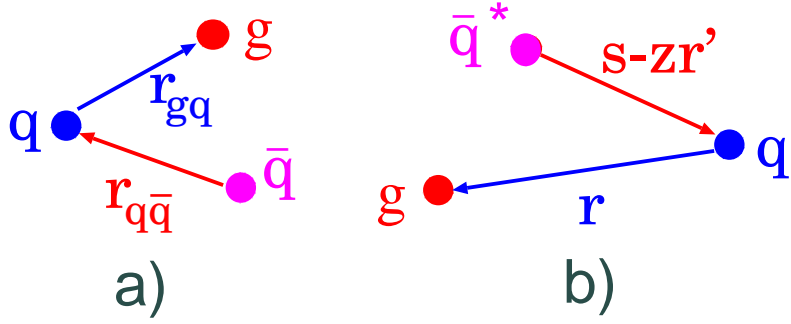}
\caption{The color dipole structure of (a) the generic quark-antiquark-gluon 
system of dipoles and (b) of the $\bar{q}^*qg$ system which emerges in the
 $\textsf{S}$-matrix structure of the two-body density
matrix for excitation $q^*\to qg$.}
\label{fig:QuarkAntiquarkGluonDipole}
\end{center}
\end{figure}
Here quark and gluon couple to the color triplet.
The dipole cross section for the color singlet $\bar{q}qg$ state
has been derived in \cite{NZ94}, the $\textsf{S}$-matrix derivation 
with the quark-antiquark basis description of the gluon is found 
in Appendix A of ref. \cite{SingleJet}.
For the generic 3-body state
shown in Fig.~\ref{fig:QuarkAntiquarkGluonDipole} it equals
\bea
\sigma_3(\br_{q\bar{q}},\br_{gq})= {C_A\over 2 C_F}\bigl[\sigma(\br_{gq})+
\sigma(\br_{g\bar{q}})
-\sigma(\br_{q\bar{q}})\bigr]+\sigma(\br_{q\bar{q}})\,,
\label{eq:3.2.1} 
\eea 
where $\br_{g\bar{q}}= \br_{gq}+\br_{q\bar{q}}$.
The configuration of color dipoles for the case of our interest
is shown in Fig.~\ref{fig:QuarkAntiquarkGluonDipole} (see the related derivation
in \cite{NPZcharm}). For the $\bar{q}^*qg$ state the relevant
dipole sizes in (\ref{eq:3.2.1}) equal 
\bea
\br_{q\bar{q}}&=&\bb_q-\bb'=\bs-z\br\,,\nonumber\\
\br_{gq}&=&\bb_{g}-\bb_q=\br\,,\nonumber\\
\br_{g\bar{q}}&=&\bb_g-\bb'=\bs+\br-z\br'\,,\
\label{eq:3.2.2} 
\eea 
whereas for the $q^*\bar{q}g'$ state we have
\bea
\br_{q\bar{q}}&=&\bb-\bb_{q}'=\bs+z\br, \,,\nonumber\\
\br_{gq}&=&\bb_{g}'-\bb_q'=\br'\,,\nonumber\\
\br_{g\bar{q}}&=&\bb_{g}-\bb=\bs+z\br -\br'\,,
\label{eq:3.2.3} 
\eea 
so that
\bea
&&\sigma_{\bar{q}^*qg}={C_A\over 2 C_F}\bigl[\sigma(\br)+
\sigma(\bs+\br-z\br')
-\sigma(\bs-z\br')\bigr]+\sigma(\bs-z\br)\,, \nonumber\\
&&\sigma_{q^*\bar{q}g'}={C_A\over 2 C_F}\bigl[\sigma(-\br')+
\sigma(\bs-\br'+z\br)
-\sigma(\bs+z\br)\bigr]+\sigma(\bs+z\br)\,.
\label{eq:3.2.4} 
\eea 
The overall color-singlet $q\bar{q}g$ state has a unique color structure
and  its elastic scattering on a nucleus
is a single-channel problem. Consequently, the nuclear $\textsf{S}$-matrix
is given by the single-channel Glauber-Gribov 
formula \cite{Glauber,Gribov}
\bea 
\textsf{S}^{(3)}_{\bar{q}'g'q^*}(\bb,\bb_q',\bb_g')&=&\textsf{S}[\bb,\sigma_{q^*\bar{q}'g'}]\,,\nonumber\\
\textsf{S}^{(3)}_{\bar{q^*}qg}(\bb',\bb_q,\bb_g)&=&\textsf{S}[\bb,\sigma_{\bar{q}^*qg}]\,.
\label{eq:3.2.5} 
\eea

%%%%%%%%%%%%%%%%%  Section 4
\section{Coupled-channel $\textsf{S}$-matrix for the 4-parton state}

%%%%%%%%%%%%  section 4.1
\subsection{The basis of color-singlet $(q\bar{q}gg')$ states}

The $4$-parton $\textsf{S}$-matrix describes transitions between  color-singlet 
$(q\bar{q}gg')$ states.  It is convenient to decompose the
the $\ket{qg}$ state into the $\ket{3}, \quad \ket{6}$ and $\ket{15}$ states
and their $SU(N_c)$ generalizations (our reference to the
triplet, sextet and 15-plet states at arbitrary $N_c$ should not 
cause any confusion). Then the basis of 
color-singlet states  $\ket{q\bar{q}gg'}$  will consist of the 
$\ket{3\bar{3}}, \quad \ket{6\bar{6}}$ and $\ket{15\,\overline{15}}$
systems of color dipoles and the intranuclear evolution in the elastic
scattering of the 4-parton state off the nucleus is the three-channel problem.
The evolution starts from the $\ket{3\bar{3}}$ state what is evident
from Fig. \ref{fig:QuarkGluonDijetDensityMatrix}.
Technically, once the 
$3\times 3$ matrix $\hat{\Sigma}$ of 4-body
dipole cross sections is known, the corresponding nuclear
$\textsf{S}$-matrix will be given by the Glauber-Gribov formula
\bea 
\textsf{S}^{(4)}_{\bar{q}'g' g q}(\bb_q',\bb_g',\bb_q,\bb_g)=\textsf{S}[\bb,\hat{\Sigma}]\, . 
\label{eq:4.1.1} 
\eea 
Our immediate task is a calculation of the coupled-channel operator $\hat{\Sigma}$.

We chose a description of the gluon in the quark-antiquark
basis:
\beq
g^i_k = \bar{a}^ia_k - {1\over N_c}(\bar{a}a)\delta^i_k\, .
\label{eq:4.1.2}
\eeq
In the calculation of the $\textsf{S}$-matrices both the quark $a$ and 
the antiquark $\bar{a}$ must be considered as propagating 
at the same impact parameter. 
The generic quark-gluon state is described by a tensor
\beq
v^i_{kl} = g^i_k c_l =  \bar{a}^ia_k c_l - {1\over N_c}(\bar{a}a) c_l \delta^i_k\, .
\label{eq:4.1.3}
\eeq
There is  a unique color-triplet quark-gluon state (the normalization
of the states will be defined at the level of the $\ket{3\bar{3}},  \ket{6\bar{6}}$ 
and $\ket{15\,\overline{15}}$ systems of color dipoles) 
\beq
t_k =(\bar{a}c)a_k - {1\over N_c}(\bar{a}a)c_k
\label{eq:4.1.4}
\eeq
The sextet state is described by the traceless tensor antisymmetric in $(k,l)$,
\beq
A^i_{kl} = \bar{a}^i  (a_k c_l - a_l c_k) +
{1\over N_c -1} [(\bar{a} c)a_l - (\bar{a} a) c_l]\delta^i_k
-
{1\over N_c -1} [(\bar{a} c)a_k - (\bar{a} a) c_k]\delta^i_l\, ,
\label{eq:4.1.5}
\eeq
while the 15-plet is described by the traceless symmetric tensor
\beq
S^i_{kl} = \bar{a}^i  (a_k c_l + a_l c_k) -
{1\over N_c +1} [(\bar{a} c)a_l + (\bar{a} a) c_l]\delta^i_k
-
{1\over N_c +1} [(\bar{a} c)a_k + (\bar{a} a) c_k]\delta^i_l\, .
\label{eq:4.1.6}
\eeq

\begin{figure}[!t]
\begin{center}
\includegraphics[width = 5.5cm,angle=270]{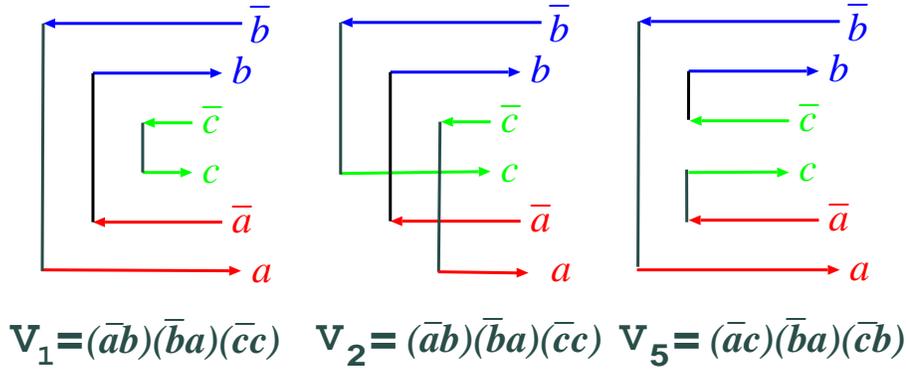}
\caption{Examples of the 6-body vertices (projection operators) 
which emerge in expansions of the $qg\bar{q}g$ states in the quark-antiquark
basis.}
\label{fig:6-bodyVertices}
\end{center}
\end{figure}

The quark, antiquark and two gluons in the color-singlet $(q\bar{q}gg')$
system of dipoles
all propagate at different impact parameters. To avoid a confusion, the gluon
in the complex conjugated state will be described by the tensor
\beq
(g')^i_k = \bar{b}^i b_k - {1\over N_c}(\bar{b}b)\delta^i_k\, ,
\label{eq:4.1.7}
\eeq
and the antitriplet state is
\beq
\bar{t}^k =(\bar{c}b)\bar{b}^k - {1\over N_c}(\bar{b}b)\bar{c}^k\, .
\label{eq:4.1.8}
\eeq
The overall color-singlet 
$\ket{3\bar{3}},~\ket{6\bar{6}}$
and $\ket{15\,\overline{15}}$ states will be decomposed into six 6-body 
color-singlet states. The corresponding 6-body vertices (projection
operators) equal  
\bea
V_1&=& (\bar{a}b)(\bar{b}a)(\bar{c}c),~~
V_2= (\bar{a}b)(\bar{b}c)(\bar{c}a),~~
V_3= (\bar{a}a)(\bar{b}c)(\bar{c}b),  \nonumber\\
V_4&=& (\bar{a}c)(\bar{b}b)(\bar{c}a),~~
V_5= (\bar{a}c)(\bar{b}a)(\bar{c}b),~~
V_6= (\bar{a}a)(\bar{b}b)(\bar{c}c).  
\label{eq:4.1.9}
\eea
some of which are pictorially represented in Fig. \ref{fig:6-bodyVertices}.
For instance, the normalized color-singlet triplet-antitriplet state will be
\bea
\ket{3\bar{3}} = \Bigl[ -{1\over N_c} V_3
-{1\over N_c} V_4 +
 V_5 +{1\over N_c^2} V_6\Bigr] \cdot{\sqrt{N_c} \over (N_c^2-1)}\, .
\label{eq:4.1.10}
\eea
Similarly, one finds
 \bea
\ket{6\bar{6}} = \Bigl[ V_1 -  
V_2
+{1\over N_c-1}\bigl(V_3+V_4-V_5-V_6\bigr)\Bigr] {1\over \sqrt{2N_c(N_c+1)(N_c-2)}}
\, .
\label{eq:4.1.11}
\eea
\bea
\ket{15\,\overline{15}} &=& \Bigl[ V_1 +  
V_2
-{1\over N_c+1}\bigl(V_3+V_4+V_5+V_6\bigr)\Bigr]{1\over \sqrt{2N_c(N_c-1)(N_c+2)}}\, .
\label{eq:4.1.12}
\eea
These states are normalized to unity, 
$\langle{3\bar{3}}\ket{3\bar{3}} = \langle{6\bar{6}}\ket{6\bar{6}}=\langle{15\,\overline{15}}\ket{15\,\overline{15}}=1$,
the normalization coefficients are readily derived using the color-flow 
diagram technique described in Sec. IV-C below.
The diagonal and off-diagonal matrix elements of the 4-body cross
section operator in the basis of  $\ket{3\bar{3}}, \ket{6\bar{6}}$ 
and $\ket{15\,\overline{15}}$ of color dipole states will be
decomposed in terms of the matrix elements 
\beq
\sigma_{ik}=\bra{V_i}\sigma\ket{V_k}
\label{eq:4.1.13}
\eeq
with the coefficients which are readily read from the
expansions (\ref{eq:4.1.10})-(\ref{eq:4.1.12}).

Note, that each of the $\sigma_{ik}$ is a matrix element between 
the overall color-singlet 6-body configurations composed of the three
color-singlet quark-antiquark pairs. As such all of them are 
infrared-safe quantities. 

%%%%%%%%%%  SEction 4.2

%%%%%%%%%%%%%%%%%%%%%%%%%%%%%%%%%%%%%

\subsection{Projection onto the final states}

In the case of the inclusive dijet spectrum with summation
over all colors of final state quarks and gluons 
the projection onto the final state is of the form (see the
discussion in \cite{Nonlinear})
\bea
&&\sum_X\bra{X}=\sum_{R}\sqrt{\dim(R)}\bra{R\bar{R}}=\nonumber\\
&=&\sqrt{N_c}\bra{3\bar{3}}+ \sqrt{{1\over 2} N_c(N_c+1)(N_c-2)}\bra{6\bar{6}}
+\sqrt{{1\over 2} N_c(N_c-1)(N_c+2)}\bra{15\,\overline{15}}\,,
\label{eq:4.2.1}
\eea
where $\dim(R)$ is the dimension of the corresponding representation.
The averaging over the colors of the initial quark $q^*$ amounts to
taking 
\beq
\ket{in} = \ket{3\bar{3}}\cdot {1\over \sqrt{\dim(3)}} \, .
\label{eq:4.2.2}
\eeq
Then the calculation of the inclusive dijet cross section requires the
evaluation of the combination of matrix elements
\bea
&&\sum_X\bra{X} \textsf{S}[\bb,\hat{\Sigma}]\ket{in}=\bra{3\bar{3}}\textsf{S}[\bb,\hat{\Sigma}]\ket{3\bar{3}}+\nonumber\\
&+&\sqrt{{\dim(6)\over \dim{(3)}} }\bra{6\bar{6}}\textsf{S}[\bb,\hat{\Sigma}]\ket{3\bar{3}} +
\sqrt{{\dim(15)\over \dim{(3)}} }\bra{15\,\overline{15}}\textsf{S}[\bb,\hat{\Sigma}]\ket{3\bar{3}}
\label{eq:4.2.3}
\eea
Besides the inclusive spectrum one can readily consider the excitation of 
quark-gluon dijets in specific color representations. We reiterate that
they also will be infrared-safe observables.

%%%%%%% section 4.3

\subsection{Color-flow diagrams}

\begin{figure}[!t]
\begin{center}
\includegraphics[width = 4.5cm,angle=270]{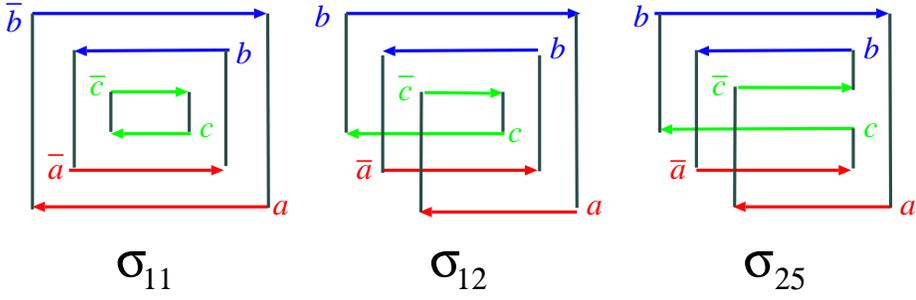}
\caption{Examples of color flow diagrams for the calculation of the
components of the 6-body dipole cross sections. The horizontal quark 
lines are multiplied  by the quark $\textsf{S}$-matrix $\textsf{S}(\bb_i)$
taken at the appropriate impact parameter, while each horizontal
antiquark line
is  multiplied by  $\textsf{S}^\dagger(\bb_i)$, the trace is taken for
each closed loop.}
\label{fig:6BodySigma}
\end{center}
\end{figure}

The calculation of the matrix elements (\ref{eq:4.1.13}) 
is greatly simplified by the technique of color-flow diagrams. 
Each matrix
element (\ref{eq:4.1.13}) corresponds to a certain 
color flow diagram. Altogether there are 21 different
color flow diagrams, the three selected cases are shown 
in Fig. \ref{fig:6BodySigma}. The number of closed loops
varies from three to one. In the calculation of the 
$\textsf{S}$-matrix elements,
\bea
\textsf{S}_{ik}=\bra{V_i}\textsf{S} \ket{V_k}\,,
\label{eq:4.3.1}
\eea
each horizontal quark line is multiplied by the quark $\textsf{S}$-matrix $\textsf{S}(\bb_i)$
taken at the appropriate impact parameter $\bb_i$, while the antiquark line
is  multiplied by  $\textsf{S}^\dagger(\bb_i)$. The trace of the product
of $\textsf{S}$-matrices is calculated for each closed loop.   

The first application of the color-flow diagrams is to
the derivation of the normalization 
factors in expansions (\ref{eq:4.1.11}). They are obtained by
assigning the factor $N_c$ to each and every loop.

Now we present the results for the three matrix elements shown
in Fig.  \ref{fig:6BodySigma}. For the sake of brevity the 
impact parameters of quarks and
antiquarks will be denoted by their symbols. One readily
finds
\bea
\textsf{S}_{11 }=\bra{V_1} \textsf{S} \ket{V_1} &=& {\rm Tr} \Bigl[ \textsf{S}(a)\textsf{S}^\dagger (\bar{b})\Bigr]~ 
{\rm Tr}  \Bigl[\textsf{S}(b)\textsf{S}^\dagger (\bar{a}) \Bigr]~{\rm Tr}  \Bigl[\textsf{S}(c)\textsf{S}^\dagger (\bar{c})\Bigr]\nonumber\\
&=&N_c^3[1-\Gamma(a-\bar{b})][1-\Gamma(b-\bar{a})][1-\Gamma(c-\bar{c})]\,. 
\label{eq:4.3.2}
\eea
The multibody $\textsf{S}$-matrices must be evaluated up to the terms quadratic
in the QCD eikonal, i.e., to the terms linear in the triplet-antitritplet
color-dipole  profile function
$\Gamma$, and the corresponding matrix element of the dipole cross section
equals
\bea
\sigma_{11}=\bra{V_1} \sigma_4\ket{V_1} &=& N_c^3\bigl[\sigma(a-\bar{b})+
\sigma(b-\bar{a})+\sigma(c-\bar{c})\bigr] \nonumber\\
&=&N_c^3\bigl[2\sigma(a-b)+\sigma(c-\bar{c})\bigr] \,.
\label{eq:4.3.3}
\eea
Each quark-antiquark loop gives the corresponding dipole cross section,
times $N_c$ to the power equal to the number of loops.
Here we took into account that the quark $a$ and antiquark $\bar{a}$,
and $b$ and $\bar{b}$ as well, propagate pairwise at equal impact parameters.

The case of $\sigma_{12}$ is a bit more complicated. Here $\textsf{S}_{12}$
is a product of two traces:
\bea
\textsf{S}_{12 }&=&\bra{V_1} \textsf{S} \ket{V_1} = 
{\rm Tr} \Bigl[\textsf{S}(b)\textsf{S}^\dagger (\bar{a}) \Bigr]~{\rm Tr} 
\Bigl[\textsf{S}(a)\textsf{S}^\dagger (\bar{b}) \textsf{S}(c)\textsf{S}^\dagger (\bar{c})\Bigr]\nonumber\\
&=&N_c[1-\Gamma(b-\bar{a})]{\rm Tr} \Bigl[ 
\textsf{S}(a)\textsf{S}^\dagger (\bar{b}) \textsf{S}(c)\textsf{S}^\dagger (\bar{c}) \Bigr]\,. 
\label{eq:4.3.4}
\eea
\begin{figure}[!t]
\begin{center}
\includegraphics[width = 10.5cm]{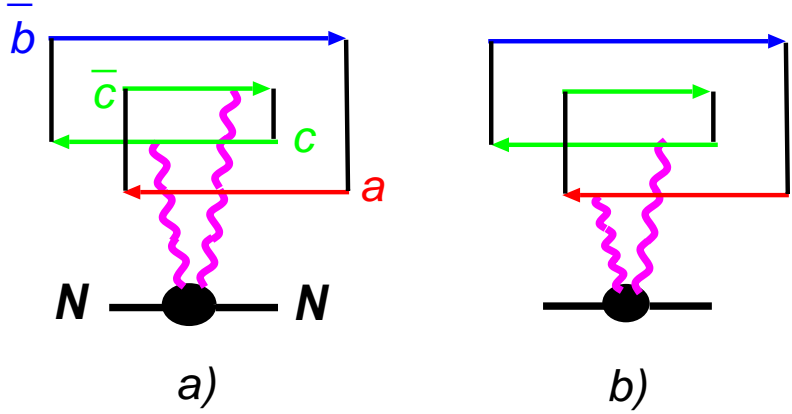}
\caption{Examples of interaction with the target
nucleon of the (a) quark-antiquark and
diquark dipole in the  $\bar{b}c\bar{c}a$ state}
\label{fig:SigmaDipoleDiquark}
\end{center}
\end{figure}

The latter trace ${\rm Tr} \Bigl[\textsf{S}(a)\textsf{S}^\dagger (\bar{b}) 
\textsf{S}(c)\textsf{S}^\dagger (\bar{c})\Bigr]$
was already encountered in our derivation of the $4$-parton $\textsf{S}$-matrix for the
production of dijets in DIS \cite{Nonlinear}. The corresponding color-flow
diagram is shown in Fig. \ref{fig:SigmaDipoleDiquark}. Here one needs to sum the
contributions to the $\bar{b}c\bar{c}a$ scattering amplitude 
from the exchange by the 2-gluon pomerons  in the $t$-channel.
The familiar diagram of Fig. \ref{fig:SigmaDipoleDiquark}a 
gives a contribution $-\chi(c)\chi(\bar{c}){\rm Tr} \Bigl(T^aT^a \Bigr)$.
The new case is when the two gluons are attached to the diquark 
$ac$ as shown in Fig. \ref{fig:SigmaDipoleDiquark}b. 
Straightforward  color algebra 
shows that the corresponding contribution to the profile function
equals
$\chi(a)\chi(c){\rm Tr}\Bigl(T^aT^a \Bigr)$. This gives rise to a simple rule:
each quark-antiquark pair, $a\bar{b}$, $a\bar{c}$, $c\bar{b}$ and $c\bar{c}$, 
contributes the corresponding triplet-antitriplet 
dipole cross section, whereas the diquark $ac$
and the anti-diquark $\bar{b}\bar{c}$ contribute the 
triplet-antitriplet dipole cross section taken with the
negative sign. The color traces give a factor $N_c$ per each
loop, one of these factors has already been put in evidence in
Eq. (\ref{eq:4.3.4}). The final result is
\bea
\sigma_{12}&=&\bra{V_1} \sigma_4\ket{V_1} = 
N_c^2\bigl[\sigma(b-\bar{a})+\sigma(a-\bar{b})
\nonumber\\
&-&\sigma(a-c)+\sigma(a-\bar{c})+
\sigma(c-\bar{b})-\sigma(\bar{b}-\bar{c}))+\sigma(c-\bar{c})\bigr]\,.
\label{eq:4.3.5}
\eea
Application of the same technique to $\sigma_{25}$ gives
\bea
\textsf{S}_{25}&=&\bra{V_1} \textsf{S} \ket{V_1} = 
{\rm Tr}\bigl[\textsf{S}(a) \textsf{S}^\dagger (\bar{a}) \textsf{S}^\dagger (\bar{c}) \textsf{S}(b) 
\textsf{S}^\dagger (\bar{a}) \textsf{S}(c)\textsf{S}^\dagger (\bar{b})\Bigr]
\label{eq:4.3.6}
\eea
with the cross section
\bea
&&\sigma_{25}=\bra{V_2} \sigma_4\ket{V_5} =
N_c\bigl[\sigma(a-\bar{c})- \sigma(a-b)\nonumber\\
&&+\sigma(a-\bar{a})-\sigma(a-c)+\sigma(a-\bar{b}) +
\sigma(b-\bar{c})-\sigma(\bar{a}-\bar{c})
\nonumber\\
&&+\sigma(c-\bar{c})-\sigma(\bar{c}-\bar{b}) +
\sigma(b-\bar{a})-\sigma(b-c)
+\sigma(b-\bar{b})\nonumber\\
&&+\sigma(c-\bar{a})-\sigma(\bar{a}-\bar{b})+
\sigma(c-\bar{b})\bigr] = N_c\sigma(c-\bar{c})
\label{eq:4.3.7}
\eea
Here we used the obvious properties $\sigma(a-\bar{a})=\sigma(b-\bar{b})=0$ and
cancellations due to equalities of the form $\sigma(c-a)=\sigma(c-\bar{a})$.
For the sake of completeness, we cite all the remaining $\sigma_{ik}$:
\bea
\sigma_{13} &=& N_c\sigma(c-\bar{c})\,,\nonumber\\
\sigma_{14} &=& N_c\sigma(c-\bar{c})\,,\nonumber\\
\sigma_{15} &=& N_c^2\bigl[2\sigma(a-b)+\sigma(c-\bar{c})+\nonumber\\
&+&\sigma(a-c)+\sigma(b-c)-\sigma(b-c)-\sigma(a-\bar{c})\bigr]\,,\nonumber\\
\sigma_{16} &=& N_c^2\sigma(c-\bar{c})\,,\nonumber\\
\sigma_{22} &=& N_c^3\bigl[\sigma(a-b)+\sigma(b-c)+\sigma(a-\bar{c})\bigr]\,,\nonumber\\
\sigma_{23} &=& N_c^2\bigl[\sigma(b-c)+\sigma(b-\bar{c})\bigr]\,,\nonumber\\
\sigma_{24} &=& N_c^2\bigl[\sigma(a-c)+\sigma(a-\bar{c})\bigr]\,,\nonumber\\
\sigma_{26} &=&  N_c\sigma(c-\bar{c})\,,\nonumber\\
\sigma_{33} &=& N_c^3\bigl[\sigma(b-c)+\sigma(b-\bar{c})\bigr]\,,\nonumber\\
\sigma_{34} &=&  N_c\sigma(c-\bar{c})\,,\nonumber\\
\sigma_{35} &=& N_c^2\bigl[\sigma(b-c)+\sigma(b-\bar{c})\bigr]\,,\nonumber\\
\sigma_{36} &=& N_c^2\sigma(c-\bar{c})\,,\nonumber\\
\sigma_{44} &=& N_c^3\bigl[\sigma(a-c)+\sigma(a-\bar{c})\bigr]\,,\nonumber\\
\sigma_{45} &=& N_c^2\bigl[\sigma(a-c)+\sigma(a-\bar{c})\bigr]\,,\nonumber\\
\sigma_{46} &=& N_c^2\sigma(c-\bar{c})\,,\nonumber\\
\sigma_{55} &=& N_c^3\bigl[\sigma(a-b)+\sigma(a-c)+\sigma(b-\bar{c})\bigr]\,,\nonumber\\
\sigma_{56} &=& N_c\sigma(c-\bar{c})\,,\nonumber\\
\sigma_{66} &=& N_c^3\sigma(c-\bar{c})\,.
\label{eq:4.3.8}
\eea

%%%%%%%   section 4.4

\subsection{The $3\times 3$ matrix of  4-parton dipole cross sections 
$\hat{\Sigma}$.}

A simple algebra gives the following 
$3\times 3$ matrix $\hat{\Sigma}$ of 4-body
dipole cross sections (we go back to the dipole parameters defined in Sect. 2):
\bea
\Sigma_{11} &=&\bra{3\bar{3}} \sigma_4 \ket{3\bar{3}} = 
{C_A\over 2C_F}\bigl[\sigma(\bs -\br'+\br)+\sigma(\br)+\sigma(\br')\bigr]\nonumber\\
&-& {1 \over N_c^2-1 }\sigma(\bs) -{C_A\over 2C_F}\cdot {1 \over N_c^2-1 }
 \Omega \,,
\label{eq:4.4.1}
\eea
where 
\bea
\Omega= 
\sigma(\bs -\br')+\sigma(\bs +\br)-\sigma(\bs -\br'+\br)-\sigma(\bs) \,.
\label{eq:4.4.2}
\eea
Similar calculation gives
\bea
\Sigma_{22}=\bra{6\bar{6}} \sigma_4 \ket{6\bar{6}} & =&
{3N_c+1 \over N_c+1}\cdot {1\over 2}\cdot \bigl[
\sigma(\bs -\br'+\br)+\sigma(\bs)\bigr]\nonumber\\
&+&      { N_c^2+1 \over 2( N_c^2-1)}      \cdot 
\bigl[\sigma(\bs -\br'+\br)-\sigma(\bs)\bigr]\nonumber\\
&+&{N_c\over N_c^2-1}\bigl[\sigma(\br)+\sigma(\br')\bigr]
\nonumber\\
&-&  {N_c \over 2(N_c+1)}\cdot 
\left[1+{1 \over (N_c-1)^2}\right] \Omega\,,\nonumber\\
\label{eq:4.4.3}
\eea
\bea
\Sigma_{33}=\bra{15\overline{15}} \sigma_4 \ket{15\overline{15}} & =&
{3N_c-1 \over N_c-1}\cdot {1\over 2}\cdot \bigl[
\sigma(\bs -\br'+\br)+\sigma(\bs)\bigr]\nonumber\\
&+&      {N_c^2+1 \over 2(N_c^2-1)}      \cdot 
\bigl[\sigma(\bs -\br'+\br)-\sigma(\bs)\bigr]\nonumber\\
&-&{N_c\over N_c^2-1}\bigl[\sigma(\br)+\sigma(\br')\bigr]
\\\label{eq:4.4.4}
&-&  {N_c \over 2(N_c-1)}\cdot 
\left[1 + {1 \over (N_c+1)^2}\right] \Omega\,.\nonumber
\eea
All the off-diagonal matrix elements for  transition  between different 
color representations are proportional to $\Omega$:
\bea
\Sigma_{21}=\bra{6\bar{6}} \sigma_4 \ket{3\bar{3}} =
-{N_c^2 \over (N_c-1)(N_c^2-1)}\sqrt{{N_c-2\over 2(N_c+1)}}
\Omega\,,
\label{eq:4.4.5}
\eea
\bea
\Sigma_{31}=\bra{15 \,\overline{15}} \sigma_4 \ket{3\bar{3}} =
-{N_c^2 \over (N_c+1)(N_c^2-1)}\sqrt{{N_c+2\over 2(N_c-1)}}
\Omega\,,
\label{eq:4.4.6}
\eea
\bea
\Sigma_{32}=\bra{15 \,\overline{15}} \sigma_4 \ket{6\bar{6}} =
-{1\over 2}\cdot {N_c^2 \over (N_c^2-1)}
\sqrt{{N_c^2-4\over N_c^2-1}}
\Omega\,.
\label{eq:4.4.7}
\eea
Note, that the off-diagonal $\Omega$ has 
precisely the same color-dipole structure as the off-diagonal
$\sigma_{18}$ which describes the excitation
$q\bar{q}$ dipole from the color-singlet to  color-octet state 
\cite{Nonlinear}. This off-diagonal matrix element vanishes
if either $\br=0$ of $\br'=0$, when the pointlike
$\ket{qg}$ and $\ket{q'g'}$ Fock states cannot be resolved.

%%%%%%%%%%%%  section 4.5
\subsection{The pointlike triplet, sextet and 15-plet dipoles
and Casimir operators}

In the limit of $\br=\br'=0$, the 4-body states reduce to the
pointlike triplet-antitriplet, sextet-antisextet and 15-$\overline{15}$ dipoles.

Indeed, in this limit
\beq
\Sigma_{11}=\sigma(\bs)
\label{eq:4.5.1}
\eeq
as expected, while
\bea
\Sigma_{22}&=&{3N_c+1 \over N_c+1}\sigma(\bs)\,,\nonumber\\
\Sigma_{33}&=&{3N_c-1 \over N_c-1}\sigma(\bs)\,.
\label{eq:4.5.2}
\eea
The Feynman diagrams of Fig.~\ref{fig:SigmaDipoleFeynman} make it obvious that
for partons in the representation $R$
the dipole cross section must be proportional to the Casimir 
operator $C_R$. Consequently, the coefficients in (\ref{eq:4.5.2}) must equal 
the ratio $C_R/C_F$ (a factor $C_F$
for the triplet-antitriplet color dipole had been absorbed into the definition
of $\sigma(\bs)$, see Eq.~(\ref{eq:3.1.3})). The derivation of $C_R$ by the color-flow diagram
technique proceeds as follows:

\begin{figure}[!t]
\begin{center}
\includegraphics[width = 3.5cm,angle=270]{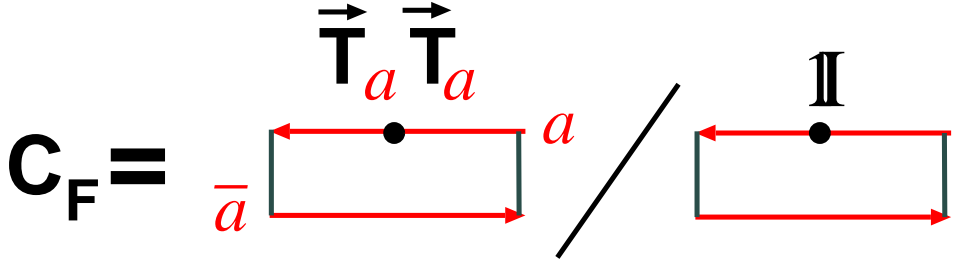}
\caption{The color-flow diagram representation of the quark Casimir
operator $C_F$.}
\label{fig:QuarkCasimir}
\end{center}
\end{figure}

We recall that the calculation of $C_F$ for the quark.\beq
C_F= {{\rm Tr}( T^aT^a) \over {\rm Tr}  \openone}\, ,
\label{eq:4.5.3}
\eeq 
can  be represented in
terms of traces of color loop diagrams as shown in Fig.~\ref{fig:QuarkCasimir}.
In order to avoid a confusion in the description of the conjugate
states, it is convenient to represent the sextet $qg$ state in terms of the
three different quark fields,  
\bea
A^i_{kl} = \bar{a}^i  (b_k c_l - b_l c_k) +
{1\over N_c -1} [(\bar{a} c)b_l - (\bar{a} b) c_l]\delta^i_k
-
{1\over N_c -1} [(\bar{a} c)b_k - (\bar{a} b) c_k]\delta^i_l\, .
\label{eq:4.5.4}
\eea
One readily finds that
\bea
\bar{A}A &\propto& (\bar{a}a)(\bar{b}b)(\bar{c}c) - 
(\bar{a}a)(\bar{b}c)(\bar{c}b)\nonumber\\
&+& {1\over N_c-1}(\bar{a}c)(\bar{b}a)(\bar{c}b)
+ {1\over N_c-1}(\bar{a}a)(\bar{b}b)(\bar{c}c)\nonumber\\
&-&{1\over N_c-1}(\bar{a}c)(\bar{b}b)(\bar{c}a)
-{1\over N_c-1}(\bar{a}b)(\bar{b}a)(\bar{c}c)\,.
\label{eq:4.5.5}
\eea
In the quark representation the Casimir operator equals
\beq
(\bT_b+\bT_c-\bT_a)^2 = 3C_F+2(\bT_b\bT_c)-2(\bT_a\bT_b)- 2(\bT_c\bT_a)\,.
\label{eq:4.5.6}
\eeq
\begin{figure}[!t]
\begin{center}
\includegraphics[width = 7.5cm,angle=270]{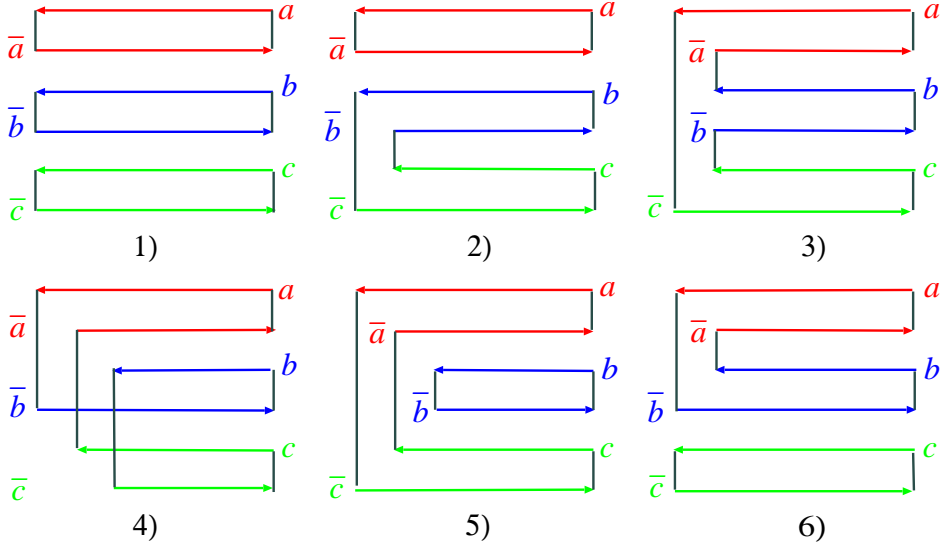}
\caption{The color-flow diagrams for the derivation of the  Casimir
operator $C_F$ for sextet and 15-plet $qg$ states in the quark-antiquark
representation.}
\label{fig:CasimirQuarkGluon}
\end{center}
\end{figure}

The six color-flow diagrams generated by the expansion
(\ref{eq:4.5.5}) are shown in Fig. \ref{fig:CasimirQuarkGluon}.
The straightforward calculation of the corresponding traces, putting
the $\bT_i$ on the relevant horizontal lines in the loops gives
\beq
C_6={3N_c+1 \over N_c+1}C_F
\label{eq:4.5.7}
\eeq
The similar expansion for the 15-plet state reads
\bea
\bar{S}S &\propto& (\bar{a}a)(\bar{b}b)(\bar{c}c) + 
(\bar{a}a)(\bar{b}c)(\bar{c}b)\nonumber\\
&+& {1\over N_c+1}(\bar{a}c)(\bar{b}a)(\bar{c}b)
+ {1\over N_c+1}(\bar{a}a)(\bar{b}b)(\bar{c}c)\nonumber\\
&+&{1\over N_c+1}(\bar{a}c)(\bar{b}b)(\bar{c}a)
+{1\over N_c+1}(\bar{a}b)(\bar{b}a)(\bar{c}c)
\label{eq:4.5.8}
\eea
and
\beq
C_{15}={3N_c-1 \over N_c-1}C_F
\label{eq:4.5.9}\,.
\eeq
This completes the check of the formulas (\ref{eq:4.5.2}). 

%%%%%%%% section 4.6
\subsection{The $N_c\to -N_c$ transformation between
the sextet and 15-plet matrix elements}

As a function of $N_c$, the Casimir operators
and matrix elements for transitions 
containing the 
sextet and 15-plet states satisfy a curious 
symmetry 
\bea
C_{15}(N_c)&=&C_6(-N_c)\,,\nonumber\\
\Sigma_{33}(N_c)&=&\sigma_{22}(-N_c)\,,\nonumber\\
\Sigma_{13}(N_c)&=&-\sigma_{12}(-N_c)\,.
\label{eq:4.6.1}
\eea
Evidently, the relative minus sign in the last line of (\ref{eq:4.6.1}) is a matter of
convention for the basis states. We do not offer any 
straightforward group-theoretic
explanation for this transformation (see, however, a discussion of
the correspondence between the symmetric and antisymmetric representations
in Cvitanovic's
lectures \cite{Predrag}).

%%%%%%%%%%%%  section 4.7
\subsection{Large-$N_c$ properties of $\Sigma$}

The application of the above derived $\hat{\Sigma}$ to the dijet
production of the free-nucleon target is straightforward. 
In the case of the nuclear target one has to solve the secular equation 
for the eigenvalues and eigenstates of $\hat{\Sigma}$. It is a cubic
equation, can be solved in radicals and the corresponding eigenfunctions
are directly calculable. The further
application of the Sylvester expansion \cite{Nonlinear} to (\ref{eq:4.1.1})
is straightforward. Unfortunately, because of the radicals 
the relevant Fourier transforms in (\ref{eq:2.12}) can only
be performed numerically.
Simple algebraic formulas for eigenvalues and analytic results 
for the dijet spectra are, however, obtained in the large-$N_c$
approximation. The higher order terms of expansion in inverse powers
of $N_c$ can also be presented in an analytic form \cite{Nonlinear}.

Note, that for large $N_c$
\bea
\Sigma_{31}&=&\Sigma_{21}={1\over N_c\sqrt{2}}\Omega\,, \nonumber\\
\Sigma_{32}&=& {1\over 2}\Omega\,, \nonumber\\
\Sigma_{33} &=& \Sigma_{22}=2\sigma(\bs-\br'+\br)+\sigma(\bs) -{1\over 2}\Omega\,,
\label{eq:4.7.1}
\eea
which shows that one must first diagonalize the matrix $\hat{\Sigma}$
in the $\ket{6\bar{6}},\ket{15\,\overline{15}}$ sector. The two
eigenvalues are 
\bea
\Sigma_{2,3}=\sigma_{22}\pm{1\over 2}\Omega
\label{eq:4.7.2}
\eea
and the corresponding eigenstates are
\bea
\ket{2}&=&{1\over \sqrt{2}}\bigl( \ket{6\bar{6}}+\ket{15\,\overline{15}}\bigr)
={V_1\over N_c^{3/2}}\,,
\nonumber\\
\ket{3}&=&{1\over \sqrt{2}}\bigl( \ket{6\bar{6}}-\ket{15\,\overline{15}}\bigr)
={V_2\over N_c^{3/2}}\,.
\label{eq:4.7.3}
\eea
In the basis of the states $\ket{1}=\ket{3\bar{3}},~\ket{2}$ and $\ket{3}$
the matrix $\hat{\Sigma}$ takes the form ($\Sigma_1=\Sigma_{11}$)
\bea
\hat{\Sigma}=
\left(\begin{array}{lll}\Sigma_1 & {1\over N_c}\Omega & 0\\
{1\over N_c}\Omega &  \Sigma_2 & 0\\
0& 0& \Sigma_3
\end{array}\right)\, ,
\label{eq:4.7.4}
\eea
where
\bea
\Sigma_1&=&\sigma(\bs +\br -\br')+\sigma(\br)+\sigma(\br')\,,\nonumber\\
\Sigma_2&=&2\sigma(\bs +\br -\br')+\sigma(\bs)= C_2\sigma(\bs +\br -\br')+\sigma(\bs)\,.
\label{eq:4.7.5}
\eea
Here we show an explicit dependence on the Casimir operator for the
large-$N_c$ eigenstate 
\beq
C_2+1 ={C_6\over C_F} ={C_{15} \over C_F} = 3\,.
\label{eq:4.7.6}
\eeq
As a matter of fact, at large $N_c$ the quark and gluon colors in the sextet and 15-plet
states become decorrelated, so that $C_6=C_{15}=C_A+C_F$ and
\beq
C_2 ={C_A\over C_F}\,.
\label{eq:4.7.7}
\eeq

To the leading order in the  $1/N_c$ expansion, the state $\ket{3}$ is not excited by
single-gluon exchange from the initial 
state $\ket{1}=\ket{3\bar{3}}$. This decoupling is obvious also from 
the projection onto the final states (\ref{eq:4.6.1}), which at large $N_c$ 
reads
\bea
\sum_X\bra{X}=\sum_{R}\sqrt{\dim(R)}\bra{R\bar{R}}
=\sqrt{N_c}\bra{1}+ (\sqrt{N_c})^3\bra{2}\,.
\label{eq:4.7.8}
\eea

In the new basis the non-Abelian intranuclear evolution of the 4-body $qg\bar{q}g'$
state becomes the two-channel problem. Expansion over the final states takes
the form
\bea
\sum_X\bra{X} \textsf{S}[\bb,\hat{\Sigma}]\ket{1}
=\sqrt{N_c}\bra{1}\exp\left[-{1\over 2}\hat{\Sigma} T\right]\ket{1}+ 
(\sqrt{N_c})^3\bra{2}\exp\left[-{1\over 2}\hat{\Sigma} T\right]\ket{1}\,.
\label{eq:4.7.9}
\eea
To the leading order in $N_c$, matrix element of the $\textsf{S}$-matrix in the first term equals
\bea
\bra{1}\exp\left[-{1\over 2}\hat{\Sigma} T\right]\ket{1}=\exp\left[-{1\over 2}\Sigma_{1} T\right]
=\exp\left\{-{1\over 2}[\sigma(\bs)+\sigma(\br)+\sigma(\br')]T\right\}\, .
\label{eq:4.7.10}
\eea
Making use of the Sylvester expansion technique used in \cite{Nonlinear}, 
for the second matrix element one finds
\bea
\bra{2}\exp\left[-{1\over 2}\hat{\Sigma} T\right]\ket{1}=
\Omega \cdot {1\over N_c}\cdot{\exp\left[-{1\over 2}\Sigma_{1} T\right]-
\exp\left[-{1\over 2}\Sigma_{2} T\right]
\over \Sigma_2-\Sigma_1} \,.
\label{eq:4.7.11}
\eea
The integral representation of Ref. \cite{Nonlinear},
\bea
{\exp\left[-{1\over 2}\Sigma_{1} T\right]-\exp\left[-{1\over 2}\Sigma_{2} T\right]
\over \Sigma_2-\Sigma_1} ={1\over 2}T \int_0^1 d\beta 
\exp\left[-{1\over 2}(\beta \Sigma_{1} +(1-\beta) \Sigma_{2}) T\right]\,,
\label{eq:4.7.12}
\eea
makes explicit a decomposition into the Initial State and Final State distortions
described by the cross sections $\Sigma_{1}$ and $\Sigma_{2}$, respectively.
Our final result for the sum over final states reads 
\bea
\sum_X\bra{X} \textsf{S}[\bb,\hat{\Sigma}]\ket{1}&=&\sum_X\bra{X}\exp\Bigl[-{1\over 2}\hat{\Sigma} T\Bigr]\ket{1}=
\nonumber\\
&&\sqrt{N_c} \Biggl\{\exp\Bigl[-{1\over 2}
[\sigma(\bs+\br-\br')+\sigma(\br)+\sigma(\br')]T\Bigr] +  \nonumber\\
&+& \Omega \cdot T \int_0^1 d\beta \exp\Bigl[-{1\over 2}(\beta \Sigma_{1} +
(1-\beta) \Sigma_{2}) T\Bigr]\Biggr\}\,.
\label{eq:4.7.13}
\eea

The systematic approach to 
perturbation $1/N_c$ expansion in has been developed in
\cite{Nonlinear} on an example of quark-antiquark dijets in DIS. Its 
extension to quark-gluon dijets is straightforward, we will not
dwell into that in this communication.
 
%%%%%%%%%%%  section 5

\section{The linear $k_{\perp }$-factorization for dijets form the free nucleon target}

The $\textsf{S}$-matrices in the master formula (\ref{eq:2.12}) depend only on the dipole
parameters $\bs,\br,\br'$. In the case of the free nucleon target one can integrate over
the overall impact parameter and represent the integrand of  Eq. (\ref{eq:2.12})
in terms of the dipole cross sections:
\bea
2\int d^2\bb \sum_{X}\bra{X}&\Biggl\{&
\textsf{S}^{(4)}_{\bar{q}g' qg}(\bb_q',\bb_g',\bb_q,\bb_g) 
+ \textsf{S}^{(2)}_{\bar{q^*}q^*}(\bb',\bb)        \nonumber\\
& -&
\textsf{S}^{(3)}_{\bar{q}g'q^*}(\bb,\bb_q',\bb_g')
- \textsf{S}^{(3)}_{\bar{q^*}qg}(\bb',\bb_q,\bb_g) \Biggr\}\ket{in}\nonumber\\
&=& \sigma_{\bar{q}^*qg}+\sigma_{q^*\bar{q}g'}-\Sigma_{11} 
+\sqrt{{\dim(6)\over \dim(3)}}\Sigma_{21} + \sqrt{{\dim(15)\over \dim(3)} }\Sigma_{31}\nonumber\\
&=& {C_A\over C_F}\bigl[
\sigma(\bs+\br-z\br') +  \sigma(\bs +z\br-\br')  
-\sigma(\bs+ \br -\br')-\sigma(\bs +z\br -z\br')\bigr]\nonumber\\
&-&{1\over N_c^2-1}\bigl[\sigma(\bs-z\br')+
\sigma(\bs +z\br )-   \sigma(\bs)  
-\sigma(\bs +z\br-z \br')\bigr]\nonumber\\
&+& {C_A\over C_F}\Omega
\label{eq:5.1}
\eea
Now we apply the $k_{\perp}$-factorizaton representation for the free-nucleon dipole cross
section. For instance, one readily finds 
\bea
\Omega = \int d^2\bkappa f(\bkappa)[1-\exp(i\bkappa\br)][1-\exp(-i\bkappa\br')]\exp(i\bkappa\bs)\,.
\label{eq:5.2} 
\eea 

The momentum space wave function of the $qg$ Fock state of the physical
quark is defined by the Fourier transform
\beq
\Psi(z,\bp)=\int d^2\br \Psi(z,\br)\exp(-i\bp\br)\, .
\label{eq:5.3}
\eeq
We discuss the cross sections averaged over the helicities of the
incident parton and summed over helicities of the final-state partons.
Then  $\Psi(z,\bp)$ would always enter in combinations of
the form \cite{NZ91,SingleJet}
\bea
|\Psi(z,\bp)-\Psi(z,\bp-\bkappa)|^2=
2N_c  \alpha_{S}\Bigl((Q^*)^2\Bigr)P_{gq}(z)\cdot
\left({\bp \over  \bp^{2}+\varepsilon^{2}} -
{\bp-\bkappa  \over  (\bp-\bkappa )^{2}+\varepsilon^{2}}\right)^2\, ,
\label{eq:5.4}
\eea
where $P_{gq}(z)$ is the familiar splitting function,
\beq
P_{gq}(z)=C_F {1+(1-z)^2 \over z} \, ,
\label{eq:5.5}
\eeq
and, neglecting the mass of the incident light quark,
\beq
\varepsilon^2 = z(1-z)(Q^*)^2 \,,
\label{eq:5.6}
\eeq
where $(Q^*)^2=(\bp^*)^2$ is the virtuality of the incident quark $q^*$, given by
the square of its transverse momentum in the projectile hadron.
If $\varepsilon^2$ is negligible small compared to $\bp^2$, then one can
use the large-$\bp$ approximation,
\beq
\left({\bp \over  \bp^{2}} -
{\bp-\bkappa  \over  (\bp-\bkappa )^{2}}\right)^2 = {\bkappa^2
\over \bp^{2}(\bp-\bkappa )^{2}} \, ,
\label{eq:5.7}
\eeq
{and it is worth to recall the emerging exact factorization 
of longitudinal and transverse momentum 
dependencies which is a well known feature of the high energy limit.}

Then the master formula for the free-nucleon cross section takes the form
\bea
&&{d \sigma_N (q^* \to qg) \over dz d^2\bp_q d^2\bp_g } =
 {1 \over 2(2 \pi)^4} \int d^2\bkappa f(\bkappa)\nonumber\\
&\times&\int d^2\bs d^2\br d^2\br'
\exp[-i\bDelta \bs- i\bp \br +i\bp \br']\exp(i\bkappa\bs)
 \Psi(z,\br) \Psi^*(z,\br') \nonumber \\
&\times& \Biggl\{{C_A\over 2C_F}
[1-\exp(i\bkappa\br)][1-\exp(-i\bkappa\br')]\nonumber\\
&+&{C_A\over 2C_F}[\exp(iz\bkappa\br)-\exp(i\bkappa\br)]
[\exp(-iz\bkappa\br')-\exp(-i\bkappa\br')]\nonumber\\
&-&{1\over N_c^2-1}[1-\exp(iz\bkappa\br)]
[1-\exp(-iz\bkappa\br')]\Biggr\}\nonumber\\
&=&{1 \over 2(2 \pi)^2} f(\bDelta)\Biggl\{ {C_A\over 2C_F}|\Psi(z,\bp)-\Psi(\bp - \bDelta)|^2
\nonumber\\
&+&
{C_A\over 2C_F}
|\Psi(z,\bp - \bDelta)-\Psi(\bp - z\bDelta)|^2-
{1\over N_c^2-1} |\Psi(z,\bp)-\Psi(\bp - z\bDelta)|^2
\Biggr\}
\label{eq:5.8}
\eea
A direct comparison shows that the dijet spectrum (\ref{eq:5.8}) is precisely the
differential form of the inclusive single gluon spectrum from the excitation
$q^*\to qg$ which was derived in \cite{SingleJet}. The reason emphasized in
\cite{Nonuniversality} is that the excitation $q^* \to qg$ proceeds via
one-gluon exchange and the acoplanarity momentum is precisely the transverse
momentum of the exchanged gluon. Remarkably, the color dipole structure of the 
integrand of the dijet cross section only differs from the one for the single-jet
spectrum by the shift of arguments of all the dipole cross sections by $\bs$.

The free-nucleon cross-section is a linear functional of the unintegrated
gluon density. Then, with certain reservations on the region of 
soft $\bDelta$, the 
acoplanarity distribution is a direct probe of $f(x,\bDelta)$.
First, on the pQCD side, the unintegrated gluon density $f(x,\bDelta)$
is well-defined only for sufficiently large momenta $\bDelta$ above
the soft scale. Second, from the practical point of view, any definition
of the jet momentum has an intrinsic uncertainty with whether the soft
hadron belongs to the jet or to the underlying soft event, so that
experimentally the acoplanarity momentum is well-defined only when
it is above the soft scale.

%%%%%%%%%%%%% section 6

\section{The nonlinear $k_{\perp }$-factorization for the dijet production off nuclei}

%%%%%%%%%%%%  section 6.1

\subsection{The color-dipole representation at large $N_c$}

The final Fourier representation for the leading term of the large-$N_c$ expansion
for the dijet cross section per unit area
in the impact parameter space  reads
\bea
{d \sigma (q^* \to qg) \over d^2\bb dz d^2\bDelta d^2\bp }& =& {1 \over (2 \pi)^4} \int
d^2\bs d^2\br d^2\br'\nonumber \\
&\times& \exp[-i\bDelta \bs- i\bp \br +i\bp \br']
\Psi(z,\br) \Psi^*(z,\br') \nonumber \\
&\Biggl\{&{1\over 2}\Omega \cdot T (\bb) \int_0^1 d\beta 
\exp\left[-{1\over 2}[\beta \Sigma_{1} +(1-\beta) \Sigma_{2}] T(\bb)\right]\nonumber\\
&+&\exp\left[-{1\over 2}[\sigma(\bs+\br-\br')+\sigma(\br)+\sigma(\br')]T(\bb)\right]\nonumber\\
&+&\exp\left[-{1\over 2} \sigma(\bs -z\br'+z\br) T(\bb)\right]\nonumber\\
&-&\exp\left[-{1\over 2}[\sigma(\br)+\sigma(\bs+\br-z\br')]T(\bb) \right]\nonumber\\
&-&\exp\left[-{1\over 2}[\sigma(\br')+\sigma(\bs-\br'+z\br)]T(\bb) \right]\Biggr\}
\label{eq:6.1.1} 
\eea 
Recall that the first term, $\propto \Omega$, describes 
the excitation from the color-triplet dipole
to sextet and 15-plet dipole states. Note, how 
the large-$N_c$ suppression of the 
off-diagonal matrix element $\Sigma_{12}$ is compensated for by a large
number of final states in the higher representations.
At large $N_c$, once the 
sextet and 15-plet states have been excited, their de-excitation back to the 
triplet state is suppressed and the further intranuclear evolution consists of the
color rotations within the higher representations. The remaining four terms 
in (\ref{eq:6.1.1}) describe the rotations within the color triplet states.

%%%%%%%%%%%%  Section 6.2

\subsection{Unintegrated collective nuclear glue and isolation of
initial state distortions}

The transformation from the color-dipole to the momentum representation
is furnished by the $k_{\perp}$-factorization formula (\ref{eq:3.1.4})
and its generalization to the nuclear target. For the latter we
adopt  the collective nuclear unintegrated gluon density 
per unit area in the impact parameter plane, $\phi(\bb,x,\bkappa)$,  
as defined  
in terms of the nuclear profile function 
\cite{NSSdijet,NSSdijetJETPLett,Nonlinear,NonlinearJETPLett}:
\bea 
\Gamma[\bb,\sigma(x,\br)] =1-\exp\left[-{1\over 2}\sigma(x,\br)T(\bb)\right]
\equiv \int d^2\bkappa
\phi(\bb,x,\bkappa) \Big[1 - \exp(i\bkappa\br) \Big] \, .
\label{eq:6.2.1} 
\eea
The utility of $\phi(\bb,x,\bkappa)$ stems from the 
observation that the driving term of small-$x$ nuclear structure 
functions, the amplitude of coherent diffractive production 
of dijets off nuclei and the single-quark
spectrum from the $\gamma^* \to q\bar{q}$ excitation off a nucleus 
all take the familiar $k_{\perp}$-factorization form in terms of
$\phi(\bb,x,\bkappa)$. The so defined collective nuclear
glue $\phi(\bb,x,\bkappa)$
satisfies the sum rule
\beq
\int d^2\bkappa \phi(\bb,x,\bkappa) = 1 - \textsf{S}[\bb,\sigma_{0}(x)] \, ,
\label{eq:6.2.2}
\eeq
where $\sigma_0(x)$ is the dipole cross section for large dipoles.
The multiple-scattering expansion of $\phi(\bb,x,\bkappa)$ in terms
of the collective glue of $j$-overlapping nucleons 
in the Lorentz-contracted nucleus
and its  nuclear shadowing and antishadowing properties are found in
\cite{NSSdijet,NSSdijetJETPLett,Nonlinear,NonlinearJETPLett} and need not
be repeated here. We only cite the formula for the so-called saturation 
scale 
\beq
Q_A^2(\bb,x)\approx {4\pi^2 \over N_c}\alpha_S(Q_A^2)G(x,Q_A^2)T(\bb) 
\label{eq:6.2.3}
\eeq
and reiterate that at a large saturation scale $\phi(\bb,x,\bkappa)$
is well-defined not only for perturbative values of $\bkappa^2$
below $Q_A^2(\bb,x)$, its continuation to the soft region is also
stable. To this end we recall that although 
$\sigma_0(x)$ enters the multiple-scattering expansion 
for $\phi(\bb,x,\bkappa)$, the final form of  $\phi(\bb,x,\bkappa)$
is exclusively controlled by $Q_A^2(\bb,x)$ and does  not depend on 
the auxiliary soft parameter $\sigma_0(x)$ \cite{Nonlinear}. 
We also note, that the nuclear profile function satisfies the 
$s$-channel unitarity
bound for the partial waves of the dipole-nucleus scattering,
$\Gamma[\bb,\sigma(x,\br)] \leq 1$, while the partial wave of 
the impulse approximation (IA) overshoots the $s$-channel  
unitarity bound for 
sufficiently heavy nucleus, $\Gamma^{(IA)}[\bb,\sigma(x,\br)]
= {1\over 2}\sigma(x,\br)T(\bb) > 1$. As such, the unintegrated 
collective nuclear gluon density $\phi(\bb,x,\bkappa)$ defined by
Eq.~(\ref{eq:6.2.1}) unitarizes 
the density of partons in a Lorentz-contracted ultrarelativistic
nucleus.

Still another convenient quantity is
\beq
\Phi(\bb,x,\bkappa)=\textsf{S}[\bb,\sigma_{0}(x)]\delta(\bkappa) +\phi(\bb,x,\bkappa)
\label{eq:6.2.4}
\eeq
in terms of which 
\beq
\exp\left[-{1\over 2}\sigma(x,\br)T(\bb)\right] = \int d^2\bkappa \Phi(\bb,x,\bkappa)
\exp(i\bkappa\br)\,.
\label{eq:6.2.5}
\eeq
We shall also encounter the collective glue for a slice $0<\beta<1$ of the nucleus:
\beq
\exp\left[-{1\over 2}\beta \sigma(x,\br)T(\bb)\right] = \int d^2\bkappa \Phi(\beta;\bb,x,\bkappa)
\exp(i\bkappa\br)\,,
\label{eq:6.2.6}
\eeq
and the intranuclear attenuation-distorted wave
function in the dipole and momentum representations,
\bea 
\Psi(\beta,x;z,\br)&=& \Psi(z,\br)\exp\left[-{1\over 2}\beta \sigma(x,\br)T(\bb)\right],\nonumber\\
\Psi(\beta,x;z,\bp)&=&\int d^2\br \Psi(\beta;z,\br)\exp(-i\bp\br)
= \int d^2\bkappa \Psi(z,\bp-\bkappa)\Phi(\beta;\bb,x,\bkappa) \, .
\label{eq:6.2.7}
\eea
Hereafter, unless it may cause a confusion, we
suppress the variable $x$ in gluon densities, dipole cross sections and distorted wave functions.

%%%%%%%%%%%%%%%  section 6.3

\subsection{Excitation of color-triplet quark-gluon dipoles}

First, we rewrite the last four terms in
the integrand of (\ref{eq:6.1.1}) in  terms of the distorted wave functions. Then we
make use of the Fourier representation (\ref{eq:6.2.5}), (\ref{eq:6.2.7}):
\bea
\left.{d \sigma \bigl(q^* \to qg\bigr) \over d^2\bb dz d^2\bDelta d^2\bp }\right|_{3}& =& {1 \over (2 \pi)^4} \int
d^2\bs d^2\br d^2\br'\exp[-i\bDelta \bs- i\bp \br +i\bp \br']\nonumber \\
&\Biggl\{&\Psi(1;z,\br) \Psi^*(1;z,\br')\exp\left[-{1\over 2}\sigma(\bs+\br-\br')T(\bb)\right]\nonumber\\
&+&\Psi(z,\br) \Psi^*(z,\br')\exp\left[-{1\over 2} \sigma(\bs -z\br'+z\br) T(\bb)\right]\nonumber\\
&-&\Psi(1;z,\br) \Psi^*(z,\br')\exp\left[-{1\over 2}\sigma(\bs+\br-z\br')T(\bb) \right]\nonumber\\
&-&\Psi(z,\br) \Psi^*(1;z,\br')\exp\left[-{1\over 2}\sigma(\bs-\br'+z\br)T(\bb) \right]\Biggr\}\nonumber\\
&=& {1 \over (2 \pi)^4} \int
d^2\bs d^2\br d^2\br'd^2\bkappa \Phi(\bb,\bkappa)\exp[-i\bDelta \bs- i\bp \br +i\bp \br']\nonumber \\
&\Biggl\{&\Psi(1;z,\br) \Psi^*(1;z,\br')\exp[i\bkappa(\bs+\br-\br')]\nonumber\\
&+&\Psi(z,\br) \Psi^*(z,\br')\exp[i\bkappa(\bs -z\br'+z\br)]\nonumber\\
&-&\Psi(1;z,\br) \Psi^*(z,\br')\exp[i\bkappa(\bs+\br-z\br')]\nonumber\\
&-&\Psi(z,\br) \Psi^*(1;z,\br')\exp[i\bkappa(\bs-\br'+z\br)]\Biggr\}\nonumber\\
&=& {1 \over (2 \pi)^2} \Phi(\bb,\bDelta)\left| \Psi(1;z,\bp-\bDelta)- \Psi(z,\bp-z\bDelta)\right|^2\nonumber\\
&=& {1 \over (2 \pi)^2}\phi(\bb,\bDelta)\left| \Psi(1;z,\bp-\bDelta)- \Psi(z,\bp-z\bDelta)\right|^2\nonumber\\
&+&{1 \over (2 \pi)^2} \delta(\bDelta)\left| \Psi(1;z,\bp)- \Psi(z,\bp)\right|^2 \textsf{S}[\bb,\sigma_{0}(x)]
\label{eq:6.3.1} 
\eea 
Now recall \cite{NSSdijet} that the amplitude of the coherent diffractive excitation 
$qA \to (qg) A$ is precisely proportional to 
\beq
 \Psi(z,\bp)- \Psi(1;z,\bp) = \int d^2\br \Psi(z,\br)\left\{1- \exp\left[-{1\over 2} \sigma(\br)T(\bb)\right]
\exp[-i\bp \br]
\right\}\,, 
\label{eq:6.3.2} 
\eeq 
so that the last term in (\ref{eq:6.3.1}) describes the coherent 
diffractive production of dijets. 
In the approximation of very large nucleus the diffractive dijets 
are produced exactly back-to-back. For finite nuclei instead of the delta-function
$\delta(\bDelta)$ one finds the sharp peak of the width $\bDelta^2 \lsim 1/R_A^2$
which is described by the form factor of the nucleus, the  details are found
in \cite{NSSdijet} and must not be repeated here.
The former term describes inelastic, incoherent production of color-triplet
$qg$ states.

%%%%%%%%%%%%%%%  section 6.4

\subsection{The contribution from sextet and 15-plet final states}

The evaluation of the contribution from the excitation
of higher color representations  in (\ref{eq:6.1.1}) proceeds as follows. First, 
we make use of the integral representation (\ref{eq:5.2}) for the off-diagonal 
cross section.
Second, keeping an explicit dependence on the Casimir operators $C_{6,15}$, we have
\bea 
&&\int_0^1 d\beta \exp\left[-{1\over 2}(\beta \Sigma_{1} +(1-\beta) \Sigma_{2}) T(\bb)\right]\nonumber\\
&=&\int_0^1 d\beta 
\exp\left\{-{1\over 2}\beta [\sigma(\bs +\br -\br')+\sigma(\br)+\sigma(\br')] T(\bb)\right\}\nonumber\\
&\times & \exp\left\{-{1\over 2}(1-\beta) [C_2\sigma(\bs +\br -\br')+\sigma(\bs)]T(\bb)\right\}\nonumber\\
&=&\int_0^1 d\beta\exp\left[-{1\over 2}\beta\sigma(\br)T(\bb)\right]\exp\left[-{1\over 2}\beta \sigma(\br')T(\bb)\right]
\nonumber\\
&\times&  \int d^2\bkappa \Phi(\beta;\bb,\bkappa_3)\exp[i\bkappa_3(\bs +\br -\br')] \nonumber\\
&\times&  \int d^2\bkappa_2 \Phi(C_2(1-\beta);\bb,\bkappa_2)\exp[i\bkappa_2(\bs +\br -\br')]\nonumber\\
&\times&  \int d^2\bkappa_1 \Phi(1-\beta;\bb,\bkappa_1)\exp[i\bkappa_1\bs]
\label{eq:6.4.1} 
\eea 
In this decomposition we keep the dipole form of the two attenuation factors 
$\textsf{S}[\bb,\beta \sigma(\br)]$ and $\textsf{S}[\bb,\beta \sigma(\br')]$.
They describe the coherent intranuclear distortion of the color-triplet quark-gluon
dipole 
before the excitation into the sextet and 15-plet representations at the depth
$\beta$ from the front face of the nucleus. The way to handle these distortion
factors has already been clarified above. Note, that in contrast to the quark-antiquark
dijet production in DIS off nuclei, both the ISI and FSI distortion factors depend
on the dipole parameter $\bs$ and explicitly contribute to the acoplanarity 
distribution. 

Combining together  (\ref{eq:5.2}), (\ref{eq:6.2.7}) and (\ref{eq:6.4.1})  we obtain
the dijet spectrum from the excitation of the sextet and 15-plet dipoles
\bea
&&\left.{d \sigma \bigl(q^* \to qg\bigr) \over d^2\bb dz d^2\bDelta d^2\bp }\right|_{6+15}
 = {1 \over (2 \pi)^2} T(\bb)\int_0^1 d\beta  \nonumber\\
&\times& \int  d^2\bkappa d^2\bkappa_1 d^2\bkappa_2 d^2\bkappa_3
\delta(\bkappa +\bkappa_1+\bkappa_2+\bkappa_3-\bDelta)\nonumber\\
&\times& f(\bkappa)
\Phi(1-\beta;\bb,\bkappa_1)\Phi(C_2(1-\beta);\bb,\bkappa_2)\Phi(\beta;\bb,\bkappa_3)\nonumber\\
&\times& \left|\Psi(\beta;z,\bp-\bkappa_2-\bkappa_3)-\Psi(\beta;z,\bp-\bkappa_2-\bkappa_3-\bkappa)\right|^2
\label{eq:6.4.2} 
\eea 
The acoplanarity momentum $\bDelta$ manifestly receives four distinct contributions
which can be classified as follows.
The excitation from the color-triplet to the sextet and 15-plet states by 
single-gluon exchange with one of the nucleons of the nucleus contributes
the transverse momentum $\bkappa$. The momentum $\bkappa_3$ comes from the
ISI of the incident quark, the FSI of the $qg$ dipole in the sextet and 
15-plet representations contributes $\bkappa_1$ and $\bkappa_2$.

The emergence of the collective nuclear glue $\Phi(C_2(1-\beta);\bb,\bkappa_2)$ in 
the integrand of (\ref{eq:6.4.2}) is not accidental. While $(1-\beta)$ is a
thickness of the slice of the nuclear matter traversed by the sextet and
15-plet $qg$ dipoles, the factor $C_2$ derives from the
Casimir operators of higher representations, see Eq.~(\ref{eq:4.7.6}). That is
one more illustration of our point \cite{Nonlinear,SingleJet} that the collective
gluon field of the nucleus cannot be described by a single density function,
it is a density matrix in the space of color representations. In the considered
large-$N_c$ approximation, $C_2=C_A/C_F$ and $\Phi((1-\beta)C_A/C_F;\bb,\bkappa_2)$
is precisely the collective nuclear glue defined in terms of the color-singlet
gluon-gluon dipole.

The ISI and FSI distortions can partly be combined taking
the convolution \cite{Nonlinear}
\bea
 \int d^2\bkappa_3 d^2\bkappa_2\Phi(C_2(1-\beta);\bb,\bkappa_2) 
\Phi(\beta;\bb,\bkappa_3)\delta(\bkappa-\bkappa_2 -\bkappa_3)
= \Phi(\beta+C_2(1-\beta);\bb,\bkappa)\,.\nonumber\\
\label{eq:6.4.3} 
\eea 
which is also obvious from the color-dipole form in (\ref{eq:6.4.1}).
%%%%%%%%%%%%%%  ssection 6.5

\subsection{Nonlinear $k_{\perp}$-factorization for dijets: the universality classes}

\subsubsection{Quark-gluon vs. quark-antiquark dijets}

After application of the convolution (\ref{eq:6.4.3}), the final result for the inclusive dijet spectrum 
takes the form 
\bea
&&\frac{(2\pi)^2d\sigma_{A}(q^*\to qg)}{ d^2\bb dz d^2\bp d^2{\bDelta}} = \frac{1}{
2} T(\bb) %\nonumber\\
\int_0^1 d \beta
\int d^2\bkappa_1 d^2\bkappa f(x,\bkappa)\nonumber\\
&&\times \Phi(1-\beta,\bb,x,\bDelta -\bkappa_1 -\bkappa)
\Phi(\beta+C_2(1-\beta),\bb,x,\bkappa_1)\nonumber\\
&&\times \Bigl|
\Psi(\beta;z,\bp -\bkappa_1) %\nonumber\\
- 
\Psi(\beta; z,\bp  -\bkappa_1-\bkappa)
\Bigr|^2\nonumber\\
&&+ \phi(\bb,x,\bDelta) \Bigl|
\Psi(1;z,\bp-\Delta) -
\Psi(z,\bp -z\bDelta)
\Bigr|^2\nonumber\\
&&+\delta(\bDelta)\textsf{S}[\bb,\sigma_{0}(x)]\Bigl|
\Psi(1;z,\bp) -
\Psi( z,\bp)
\Bigr|^2\, .
\label{eq:6.5.1}
%\vskip{-0.5cm}
\eea 
which must be compared to the large-$N_c$ version of the
free-nucleon cross section (\ref{eq:5.8}).

The free-nucleon cross-section is a linear functional of the unintegrated
gluon density.  The $k_{\perp}$-factorization properties
of the nuclear cross section are much more complicated. At this point,
it is instructive to discuss (\ref{eq:6.5.1}) in conjunction with
the quark-antiquark dijet spectrum in DIS \cite{Nonlinear} and
 gluon-nucleus collisions \cite{Nonuniversality}. The spectrum of 
dijets in DIS equals 
\bea
&&\frac{(2\pi)^2d\sigma_{A}(\gamma^*\to Q\bar{Q}) }{ d^2\bb dz d^2\bp d^2\bDelta} = \frac{1}{
2} T(\bb) 
\int_0^1 d \beta
\int d^2\bkappa_1 d^2\bkappa 
\nonumber\\
&&\times f(\bkappa)\Phi(1-\beta,\bb,\bDelta -\bkappa_1 -\bkappa)
\Phi(1-\beta,\bb,\bkappa_1)\nonumber\\
&&\times \Bigl|
\Psi(\beta;z,\bp -\bkappa_1) -%\nonumber\\ 
\Psi(\beta; z,\bp -\bkappa_1-\bkappa)
\Bigr|^2
\nonumber\\
%\times 
&&+ \delta(\bDelta)\Bigl|
\Psi(1;z,\bp) -
\Psi( z,\bp)\Bigr|^2\, .
\label{eq:6.5.2}
\eea
where the first term describes the excitation of the color-dipole
from the lower (color-singlet)  to higher (octet) representation,
whereas the second term is the contribution from coherent
diffractive excitation. 
The spectrum of the quark-antiquark dijets in $gA$ collisions
is of the form 
\bea 
&&{(2\pi)^2 d\sigma_A(g^* \to Q\bar{Q}) \over dz d^2\bp_- d^2\bb d^2\bDelta}= 
\int d^2\bkappa \Phi(1;\bb,\bkappa) \Phi(1;\bb,\bDelta-\bkappa)\nonumber\\
&&\times 
|\Psi(z,\bp_- - \bkappa) -\Psi(z,\bp_- - z\bDelta) |^2  \, .
\label{eq:6.5.3}
\eea
Now we can identify the four universality classes of the nonlinear
$k_\perp$-factorization which differ by the pattern of transitions 
between the initial and final state color multiplets. They
describe the leading transitions in the large-$N_c$ approximation,
the higher order excitation and regeneration processes result in still higher
nonlinearity in gluon densities, the examples are found in \cite{Nonlinear}.

\subsubsection{Excitation of higher color representations from partons
in the lower representations}

Excitation of color-octet states in DIS, and of sextet and 15-plet states
in $qA$ interactions, belong to this universality class. The two
reactions have much similarity. In both cases the nonlinear $k_\perp$-factorization
formulas contain the free-nucleon gluon density $f(x,\bkappa)$,
which describes the transition from the $qg$ color dipole 
from the lower - triplet for $qg$ and singlet for DIS - 
to higher - sextet and 15-plet
for $qg$ and octet in DIS -  color dipoles. In both cases, the number of states 
in higher representations is by the factor $N_c^2$ larger than in the lower
representation.
In $q\bar{q}$ excitation in DIS the corresponding contribution to the
dijet spectrum is the fifth order functional of 
gluon densities. In the $qg$ case it
is the sixth order functional of gluon densities, only after the
application of the convolution (\ref{eq:6.4.3}) it takes the form
of the fifth order functional. 
Two powers of the collective nuclear glue enter implicitly
via the coherent ISI distortions of the wave function $\Psi(\beta;z,\bp)$
in the slice of the nuclear matter before excitation
of color dipoles in the higher representation, two more
powers of the collective nuclear glue describe the ISI and FSI
broadening of the acoplanarity distribution.  

The principal difference between DIS and $qA$ interactions is in the 
nuclear thickness dependence  of the distortion factors.
Namely, the factor 
$$\Phi((1-\beta),\bb,\bDelta -\bkappa_1 -\bkappa)
\Phi((1-\beta),\bb,\bkappa_1)$$ in DIS is the symmetric function
of the nuclear gluon momenta $\bkappa_1$ and $\bkappa_2=\bDelta-\bkappa_1-\bkappa$
which flow from the nucleus to the quark and antiquark (or vice versa), respectively.
It describes equal, and uncorrelated, distortion of the 
outgoing quark and antiquark waves by pure FSI.
The independence of the two distortion factors is a feature of the
large $N_c$ approximation.
For $qg$ dijets in $qA$ collisions the distortion factor
$$\Phi(1-\beta,\bb,\bkappa_2 )
\Phi(C_2(1-\beta) +\beta,\bb,\bkappa_1)$$ 
is an asymmetric one. The first source  of the asymmetry is the non-singlet color charge
of the projectile parton. 
The second source is that the two partons in 
the final state belong to different
color representations. This is best seen from in the  overall distortion factor
in (\ref{eq:6.4.2}),  
$$ 
\Phi(\beta;\bb,\bkappa_3) \Phi(C_2(1-\beta);\bb,\bkappa_2)\Phi(1-\beta;\bb,\bkappa_1)\,,
$$
before taking the convolution (\ref{eq:6.4.3}).
The FSI distortions in the slice $(1-\beta)$ of the nucleus are given by the
two last factors, of which $ \Phi(1-\beta;\bb,\bkappa_1)$ is a broadening due to
final-state rescatterings of the quark. Because $C_2=C_A/C_F$, see
Eq.~(\ref{eq:4.7.7}), the second FSI
factor, $\Phi(C_2(1-\beta);\bb,\bkappa_2)$,
corresponds to the FSI distortion of exactly the outgoing gluon wave.
To the large-$N_c$ approximation the rescatterings of the quark and gluon 
are uncorrelated.

The coherent ISI distortion of the wave functions in DIS and $qA$ collisions is
identical. However, in $qA$ collisions this coherent distortion is accompanied by an
incoherent ISI distortions of the incident quark wave 
described by $\Phi(\beta;\bb,\bkappa_3)$. In DIS
the incoherent ISI distortions are absent because the photon is a color-singlet
particle. We can anti\-ci\-pate that gluon-nucleus collisions with 
excitation of gluon-gluon dijets in higher color representations will belong 
to this universality class.

\subsubsection{Excitation of final state dipoles in exactly the same color state
as the incident parton: coherent diffraction}

To this universality class belong the exactly back-to-back dijets.
Another experimental signature of the coherent diffraction is a retention of the
target nucleus in the ground state and large rapidity gap between
the hadronic debris of the diffractive dijet and the recoil nucleus.
It is most important for DIS where coherent diffraction 
dissociation of the photon into $q\bar{q}$ dijets makes for heavy nuclei 
$\approx 50\%$ of the total DIS rate \cite{NZZdiffr}. The origin of the coherent 
diffraction is a coherent nuclear distortion of the wave function of the
$q\bar{q}$ Fock state over the
whole thickness of the nucleus. 

In the coherent diffractive excitation of $qg$ dipoles in $qA$
collisions the
$qg$ dipole must propagate in exactly the same color state as the
incident quark. The nuclear suppression factor 
$\textsf{S}[\bb,\sigma_{0}(x)]$
has the meaning of
\beq
\textsf{S}[\bb,\sigma_{0}(x)] = \left(\textsf{S}[\bb,{1\over 2}\sigma_{0}(x)]\right)^2
\label{eq:6.5.4}
\eeq
and the factor $\textsf{S}[\bb,{1\over 2}\sigma_{0}(x)]$ in the diffractive amplitude
corresponds to the intranuclear attenuation of the quark wave with the
total cross section 
\beq
\sigma_{qN} = {1\over 2}\sigma_{0}(x)\, .
\label{eq:6.5.5}
\eeq
Coherent diffractive excitation of color-octet gluon-gluon dijets in 
gluon-nucleus collisions is expected to exhibit similar properties.

Coherent diffractive excitation of $Q\bar{Q}$ dipoles in $gA$ 
collisions is allowed,
but it is suppressed at large $N_c$ by the condition that the $Q\bar{Q}$ dipole
must propagate in exactly the same color state as the incident gluon.

\subsubsection{Incoherent excitation of final state dipoles in the 
same lower color representation
as the incident parton}

An example of this universality class is an inelastic excitation of 
color-triplet $qg$ states 
in $qA$ collisions followed
by a color excitation of the target. Here both the incident parton and dijet
belong to the fundamental, i.e., lower, representation of $SU(N_c)$. The 
intranuclear evolution of such a dipole is confined to
rotations within the color-triplet state.  
This contribution is not suppressed at
large $N_c$.
The dijet cross section for this universality class 
looks like satisfying the linear $k_{\perp}$-factorization in 
terms of $\phi(\bb,x,\bDelta)$. But this is not the case: one of 
the wave functions, $\Psi(1;z,\bp_{g})$, is coherently distorted
over the whole thickness of the nucleus, so that this 
contribution is a cubic functional of the collective nuclear glue.

We can anticipate that gluon-nucleus collisions with 
excitation of color-octet gluon-gluon dijets will belong 
to this universality class, although one has to account for 
the existence of the two, $F$-coupled and $D$-coupled, octet
states.

Although superficially it looks like a subclass of 
this universality class, the coherent diffraction
is a distinct class for the property of  the exact
back-to-back dijets and the rapidity gap between the dijet
and the recoil nucleus in the ground state.

\subsubsection{Excitation of final state dipoles 
in the same higher color representation
as the incident parton}

In the realm of QCD with gluons in the adjoint representation and 
quarks in the fundamental representation, this universality class
consists of the quark-antiquark dijets in gluon-nucleus collisions.
Only in this case the initial parton (gluon) belongs to
the higher (octet) color multiplet of the final $Q\bar{Q}$ state. 
At large $N_c$, the intranuclear evolution of $Q\bar{Q}$  
will consist of
color rotations within the
space of color-octet states. The de-excitation from the color-octet 
to color-singlet $Q\bar{Q}$ dipoles is suppressed at large $N_c$.
Consequently, the non-Abelian evolution of the
$Q\bar{Q}Q'\bar{Q}'$ state becomes the single channel problem.  
The coherent diffraction excitation, in which
the initial and final color states must be identical, is likewise 
suppressed. The emerging pattern of quadratic nonlinearity 
can be related
to the large-$N_c$ gluon behaving like the color-uncorrelated 
quark and antiquark. 

The above classification exhausts reactions caused by incident
photons, quarks and gluons. However, technically all the universality 
classes have a much broader basis. Indeed, instead of an 
incident gluon one can think of the projectile which is a 
compact lump of many partons in the highest possible color 
representation. For instance, compact diquarks in the proton can 
be viewed as sextet partons. 

\subsubsection{Is an experimental separation of events belonging to
different universality classes possible?}

We reiterate that for all the universality classes their
separate  contributions
to the dijet cross section are infrared-safe quantities.
Coherent diffraction has distinct signatures and 
the experimental separation of events from this 
universality class is not a problem. 
Production of very forward dijets in proton-nucleus
collisions  evidently tags quark-nucleus collisions. 
Production of open charm 
in the proton hemisphere of of proton-nucleus collisions 
tags gluon-nucleus collisions. Incoherent processes belonging to
different universality classes are characterized by distinct
color charge of the hard dijet and this distinction is 
well defined at the parton 
level. Translating the cross-talk between color charges in the 
dijet, the spectator partons of the 
proton and the color-excited nucleus remnant into properties 
of hadronic final states can only be done within 
nonperturbative hadronization models. As an example we cite
the impact of color reconnection 
effects on the flow of slow hadrons and 
the accuracy of the  $W^{\pm}$ mass determination in $e^+e^-$
annihilation (\cite{Khoze}, for the review see \cite{ColorFlow}).

%%%%%%%%%%%%%%%  section 6.6

\subsection{The impulse approximation}

In the impulse approximation (IA) one only has to keep the terms linear
in $T(\bb)$. The transition to the IA is best seen in the color-dipole
representation (\ref{eq:6.1.1}). Recall, that our formulas for nuclear
cross section were derived in the large-$N_c$ approximation.
Here the first term, the contribution from 
the sextet and 15-plet final states, is already linear in $T(\bb)$ and
one must put the attenuation factors equal to unity. The remaining 
four exponentials must be expanded to terms linear in $T(\bb)$. Then
one would find precisely the large-$N_c$ version of Eq.~(\ref{eq:5.1})
times $T(\bb)$. The integration over impact parameters gives 
$\int d^2\bb T(\bb) =A$.
Such a comparison does not expose the r\^ole of coherent diffraction
and we revisit the issue in the momentum
representation.

We start with the sextet and 15-plet contribution in (\ref{eq:6.5.1}).
It already contains the factor $T(\bb)$. Consequently, one
must neglect ISI distortions in the wave function,
$\Psi(\beta;\bp) \Rightarrow \Psi(\bp)$, and take 
\beq
\Phi(1-\beta,\bb,x,\bDelta -\bkappa_1 -\bkappa)
\Phi(2-\beta,\bb,x,\bkappa_1) = \delta(\bDelta -\bkappa_1 -\bkappa)
\delta(\bkappa_1)\,. 
\label{eq:6.6.1}
\eeq
This way one would recover the first term
in the rhs of Eq. (\ref{eq:5.8}). In the contribution
from the excitation of the triplet dipoles one must neglect the
distortion of the wave function and take
\beq
\phi(\bb,x,\bkappa)={1\over 2}T(\bb)f(x,\bkappa)\, .
\label{eq:6.6.2}
\eeq
The second term in the rhs of Eq. (\ref{eq:5.8}) is recovered.
Finally, according to Eq.~(\ref{eq:6.3.2}) the diffractive
amplitude starts with the term linear in $T(\bb)$. Consequently,
the coherent diffractive contribution to the dijet cross spectrum
starts with 
the terms $\propto T^2(\bb)$ and vanishes in the IA.

%%%%%%%%%%%%%%%%  section 7

\section{Nuclear broadening of the acoplanarity distribution}

The nuclear broadening of the acoplanarity distribution of hard
quark-gluon dijets from $qA$ collisions is somewhat different
from the broadening of quark-antiquark jets in DIS and now we
comment on those differences. 

%%%%%%%%%%%%   section 7.1

\subsection{Coherent diffractive contribution}

The first striking difference is in the r\^ole of the coherent 
diffractive production.
It gives exactly back-to-back dijets. In the considered
approximation of single-gluon exchange in the $t$-channel
diffractive production off the free-nucleon target vanishes.
Experimentally, at HERA energies a fraction of DIS which is
diffractive does not exceed 10\% \cite{HERAdiffraction}.
In contrast to that, in DIS off heavy nuclei a 
fraction of coherent diffraction was shown to be as large
as  $\approx 50\%$ \cite{NZZdiffr}. The
existence of coherent diffractive 
mechanism  in the quark-nucleus collisions is interesting
by itself. From the practical point of view, it 
is suppressed by nuclear absorption and is marginal.

% section 7.2

\subsection{Excitation of the color-triplet states}

Inelastic excitation of color-triplet dipoles is a specific
feature of $qA$ collisions in the sense that it has no
counterpart in DIS. 
One must compare 
\bea
 \phi(\bb,x,\bDelta) \Bigl|
\Psi(1;z,\bp-\bDelta) -
\Psi(z,\bp -z\bDelta)
\Bigr|^2
\label{eq:7.2.1}
\eea
 with its IA form
\bea
{1\over 2} T(\bb)f(\bDelta)|\Psi(z,\bp-\bDelta)-\Psi(\bp - z\bDelta)|^2\,.
\label{eq:7.2.2}
\eea

The first striking distinction is that that for the free-nucleon target 
the contribution of this process vanishes at $z\to 1$, when the incident
quark's momentum is transferred entirely to the forward gluon jet. 
For the nuclear 
target this is not the case because one of the  wave functions in 
(\ref{eq:7.2.1}) is the nuclear-distorted one. Because 
$\bp-\bDelta=-\bp_q$, it takes the form $\phi(\bb,x,\bDelta) \Bigl|
\Psi(1;z,\bp_q) - \Psi(z,\bp_q) \Bigr|^2$; as a function of the 
quark-jet momentum, it is reminiscent of the coherent diffractive 
contribution, but the acoplanarity momentum distribution is given
by  the unintegrated nuclear gluon density $\phi(\bb,x,\bDelta)$.
Hereafter we consider the case of finite $(1-z)$.

A comprehensive discussion of
nuclear  properties of the ratio
\beq
R_g(\bb,\bDelta) =  {2\phi(\bb,\bDelta) \over  T(\bb)f(\bDelta)}
\label{eq:7.2.3}
\eeq
is found in \cite{Nonlinear,SingleJet}. It is nuclear-shadowed, 
$R_g(\bb,\bDelta) < 1$, for $\bDelta^2 \lsim Q_A^2(\bb)$ and it exhibits 
antishadowing property, $R_g(\bb,\bDelta) > 1$ in a broad region 
of $\bDelta^2 \gsim Q_A^2(\bb)$ . The maximum value of
$R_g(\bb,\bDelta)$ is reached at a value of $\bDelta^2$ which is larger
than $ Q_A^2(\bb)$ by a large numerical factor.

Now we turn to distortions of the wave function.
We are interested in hard dijets. If the incident quark is a valence quark of the proton,
its transverse momentum and virtuality have the hadronic scale
and can be neglected.  For hard jets
\beq
\Psi(z,\bp) \propto {\bp \over \bp^2} 
\label{eq:7.2.4}
\eeq
and, upon averaging over the azimuthal angle $\varphi$ of the gluon momentum $\bkappa$,
\beq
\langle \Psi(z,\bp-\bkappa) \rangle_\varphi
\propto {\bp \over \bp^2}\theta(\bp^2-\bkappa^2)\,. 
\label{eq:7.2.5}
\eeq
Consequently, the wave function distortion factor 
equals
\bea
\rho_{\psi}(\bb,z,\bp)={\Psi(1;z,\bp)\over \Psi(z,\bp)}=
\int^{\bp^2}  d^2\bkappa \Phi(\bb,\bkappa)=
1- \int_{\bp^2}  d^2\bkappa \Phi(\bb,\bkappa)
\,.
\label{eq:7.2.6}
\eea
For the weakly virtual incident quark it does not depend on $z$.
For hard jets, $\bp^2 \gsim Q_a^2(\bb)$, the remaining integral 
(\ref{eq:7.2.6}) can be evaluated 
following the analysis of the Cronin effect in \cite{SingleJet}. Namely, here we
can use the leading-twist approximation,
\beq
\Phi(\bb,\bkappa) ={1\over 2}T(\bb)f(\bkappa)\, ,
\label{eq:7.2.7}
\eeq
and the definition (\ref{eq:3.1.5}) with the result
\bea
&&\delta_\psi =1-\rho_{\psi}(\bb,z,\bp)= 
\int_{\bp^2}  d^2\bkappa \Phi(\bb,x,\bkappa) \nonumber\\
&\approx&  {2\pi^2 T(\bb) \alpha_S(\bp^2) \over N_c \bp^2}\cdot
{\partial G(x, \bp^2) \over \partial \log \bp^2} 
=
{1\over 2} \cdot {Q_A^2(\bb) \over \bp^2} \cdot {\alpha_S(\bp^2) \over \alpha_S(Q_A^2) G(x,Q_A^2)}\cdot
{\partial G(x, \bp^2) \over \partial \log \bp^2}\, .
\label{eq:7.2.8}
\eea 
It is important that $\delta_\psi$ is a manifestly positive valued quantity. 
\begin{figure}[!t]
\begin{center}
\includegraphics[width = 7.0cm, height= 8.0cm,width=16.0cm]{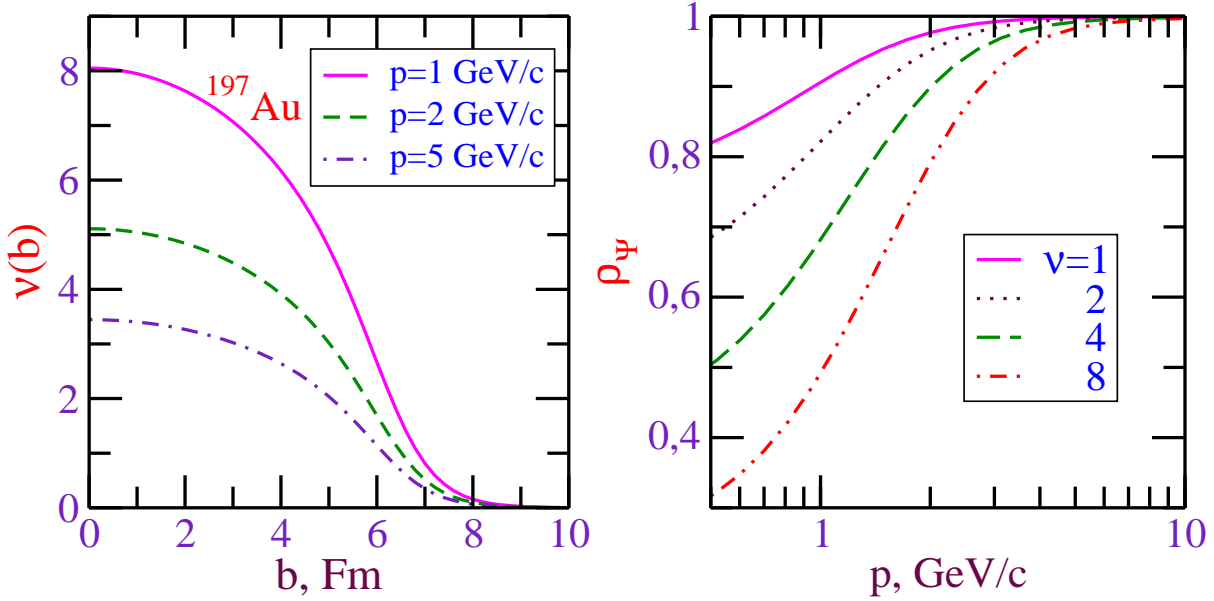}
\caption{The left panel shows the impact-parameter dependence of
the optical thickness of the gold nucleus for several values of the
gluon-jet momentum $\bp$. The momentum dependence of the
wave-function distortion factor $\rho_{\psi}(\bb,z,\bp)$ for 
several values of the optical thickness of the nucleus is
presented in the right panel.}
\label{fig:WFdistortion}
\end{center}
\end{figure}
It has a form similar to, but is numerically smaller than, 
the nuclear higher twist correction 
to $\phi(\bb,x,\bkappa)$. 

In Fig.~
\ref{fig:WFdistortion} we show the numerical results for the 
wave-function distortion factor for the gold nucleus at several
values of the optical thickness $
\nu(\bb) ={1\over 2}\sigma_0(x)T(\bb)$.
At this point one 
needs to pay a due attention to  an explicit dependence on the QCD
running coupling $\alpha_S(r)$ on the small dipole size $r$ 
in Eq.~(\ref{eq:3.1.5}). The discussion
of its impact is found in \cite{NSSdijet,Nonlinear}, in the evaluation 
of the momentum spectra this running coupling must be taken at
the largest relevant hard parameter, which in our case is $\bp^2$.
Correspondingly, in all the formulas for the dijet spectra, the dipole 
cross section for large dipoles, $\sigma_0(x)$,
must be understood as
\beq
\sigma_0(x) \Rightarrow \alpha_S(\bp^2) \cdot {4\pi^2\over N_c}\int {d\kappa^2\over \kappa^4} \cdot {\cal
F}(x,\kappa^2) = \alpha_S(\bp^2)\sigma_0(x,\infty)\,.
\label{eq:eq:7.2.9} 
\eeq
For this reason, the optical thickness of the nucleus $\nu(\bb)$ as
a function of the 
impact parameter $\bb$, shown in the left panel of Fig. \ref{fig:WFdistortion},
depends on the hard scale - the jet momentum. The wave-function
distortion factor $\rho_{\psi}(\bb,z,\bp)$ is shown in the 
right panel of  Fig. \ref{fig:WFdistortion}.
The hard regime (\ref{eq:7.2.8}) for $\delta_{\psi}$ sets in
at the momenta $p\gsim$ 1 GeV.
We reiterate that the saturated cross section $\sigma_0(x,\infty)$ is only
an auxiliary parameter which does not enter directly the observable
cross sections  - the latter only depend on the saturation scale $Q_A^2(\bb)$, 
the discussion is found in 
Ref. \cite{Nonlinear}.

In terms of the
distortion factor
$\rho_{\psi}(\bb,\bp)$ one readily finds
\bea
R_{\psi}(\bb,\bp,\bDelta)&=& { \Bigl|
\Psi(1;z,\bp-\bDelta) -
\Psi(z,\bp -z\bDelta)
\Bigr|^2 \over |\Psi(z,\bp-\bDelta)-\Psi(\bp - z\bDelta)|^2}  
\nonumber\\
&=&
{[(1-z)\bDelta -\delta_\psi (\bp - z\bDelta)]^2 \over (1-z)^2\bDelta^2}
={[(1-z)\bDelta +\delta_\psi (\bp_q - (1-z)\bDelta)]^2 \over (1-z)^2\bDelta^2}\, .
\label{eq:7.2.10}
\eea 
The overall nuclear modification factor, the ratio of the nuclear, (\ref{eq:7.2.1}), and free-nucleon, 
(\ref{eq:7.2.2}), target contributions, is a product
\bea
R_{A/N}^{(3)}(\bb,\bp,\bDelta)=R_g(\bb,\bDelta)R_{\psi}(\bb,\bp,\bDelta)
\label{eq:7.2.11}
\eea 
Here $R_g(\bb,\bDelta)$ does not depend on the jet
momentum $\bp$ except for the weak dependence through Â$\alpha_S(\bp^2)$.  
Evidently, $R_{\psi}(\bb,\bp,\bDelta)$ is
azimuthally asymmetric and favors $\bDelta$ anticollinear to the
gluon momentum and collinear to the quark momentum: 
in the back-to-back configuration, the gluon 
jet tends to have the transverse momentum smaller than the quark jet.
The dominant contribution to the nuclear dijet cross section 
comes from $\bDelta^2\sim Q_A^2(\bb)$, and for 
hard dijets the asymmetry will be
weak, of the order of $\sqrt{\delta_{\psi}} \sim Q_A(\bb)/p$. 

Alternatively, if one keeps the quark transverse momentum fixed and
increases the target mass number $A$, i.e., $Q_A^2(\bb)$ and $\delta_{\psi}$
thereof, the transverse momentum of the away gluon jet will decrease
with $A$. The form of the $q\to qg$ splitting function favors production
of the gluon jet at rapidities smaller than the quark jet. Then, the
above correlation between the acoplanarity and quark momenta shall
exhibit itself as a nuclear suppression of the away jet produced at
rapidites smaller than the rapidity of the forward trigger jet. The
numerical studies of this effect will be reported elsewhere.

%%%%%%% subsection 7.3

\subsection{Excitation of the sextet and 15-plet jets states} 

Here one must compare the contribution to the nuclear dijet spectrum (\ref{eq:6.4.2})
with its IA counterpart
\bea
\left.{ T(\bb) d \sigma_N (\bp,\bDelta) \over dz d^2\bp d^2\bDelta } \right|_{6+15}=
{1 \over 2(2 \pi)^2} T(\bb)f(\bDelta)|\Psi(z,\bp)-\Psi(\bp - \bDelta)|^2\,.
\label{eq:7.3.1}
\eea
Note, that the nuclear cross section can be cast in
the from reminiscent of a triple convolution
\bea
&&\left.{d\sigma_{A}(q^*\to qg)\over d^2\bb dz d\bp d\bDelta } \right|_{6+15}=
\nonumber\\
&=&
T(\bb)\int_0^1 d \beta
 \int  d^2\bkappa d^2\bkappa_1 d^2\bkappa_2 d^2\bkappa_3
\delta(\bkappa +\bkappa_1+\bkappa_2+\bkappa_3-\bDelta)\nonumber\\
&\times& 
\Phi(1-\beta;\bb,\bkappa_1)\Phi(C_2(1-\beta);\bb,\bkappa_2)\Phi(\beta;\bb,\bkappa_3)
\left.{ d \sigma_N (\bp-\bkappa_2-\bkappa_3,\bkappa) \over dz d^2\bp d^2\bkappa } \right|_{6+15}\, .
\label{eq:7.3.2}
\eea
which suggests that at a fixed gluon-jet momentum $\bp$, it will be a broader distribution of 
$\bDelta$ than 
the free-nucleon cross section (for the related discussion see \cite{Nonlinear}). This 
broadening is best seen for hard dijets, $\bp^2 \gg \bDelta^2,Q_A^2(\bb)$. Because the
dominant contribution comes from $\bkappa_{i}^2 \lsim Q_A^2(\bb)$, one can neglect
$ \bkappa_{2,3}$ compared to $\bp$ in the free-nucleon cross section in the integrand
of (\ref{eq:7.3.2}). Then the nuclear cross section takes the manifest convolution
form
\bea
&& \left.{d\sigma_{A}(q^*\to qg)\over  d^2\bb dz d\bp d\bDelta } \right|_{6+15}=
\nonumber\\
&=&
 %\nonumber\\
T(\bb)\int_0^1 d \beta
 \int  d^2\bkappa \Phi(1+C_2(1-\beta);\bb,\bDelta-\bkappa) 
\left.{ d \sigma_N (\bp,\bkappa) \over dz d^2\bp d^2\bkappa } \right|_{6+15}\, .
\label{eq:7.3.3}
\eea
The saturation scale for the distribution $\Phi(1+C_2(1-\beta);\bb,\bDelta-\bkappa)$ 
equals
\beq
Q_{A,eff}^2 \approx [1+C_2(1-\beta)]Q_A^2(\bb)
\label{eq:7.3.4}
\eeq
and the broadening of the acoplanarity distribution for the quark-gluon dijets is
substantially stronger than that for the quark-antiquark dijets in DIS discussed
in \cite{Nonlinear}.

%%%%  section 8
\section{The monojets from dijets: fragmentation vs. genuine dijets}

%%%%%  section 8.1
\subsection{Monojets from dijets in the free-nucleon reactions}

In the above discussion we implicitly assumed that the 
quark and gluon hard jets are separated by a large azimuthal angle 
and the acoplanarity momentum is small compared to the
jet momenta, $\bDelta^2 \lsim \bp^2, (\bp-\bDelta)^2$. The
interesting 
new situation is encountered when the quark and gluon jets 
start merging. 
Specifically, the
wave function $\Psi(z,\bp -z\bDelta)$ has a pole
when $\bp-z\bDelta=0$, i.e.,
when the gluon and quark are collinear,
\beq
\bp_g=z\bDelta,~~\bp_q=z_q\bDelta=\bDelta-\bp= (1-z)\bDelta=z_q\bDelta\, .
\label{eq:8.1.1}
\eeq
In the vicinity of the pole the $qg$ production cross section
has the factorized form
\bea
\left.{d \sigma_N (q^* \to qg) \over dz d^2\bp d^2\bDelta } \right|_{monojet}
={1 \over 2(2 \pi)^2} f(\bDelta)|\Psi(z,\bp - z\bDelta)|^2\,.
\label{eq:8.1.2}
\eea

Now recall that $\Psi(z,\bp - z\bDelta)$ is precisely a 
probability amplitude 
to find the gluon with the  momentum 
$\bk_{\perp} =\bp-z\bDelta$ transversal with respect to
the axis of the quark jet with the momentum $\bDelta$,
and  $|\Psi(z,\bp - z\bDelta)|^2$ of Eq.~(\ref{eq:5.4})
is proportional to the familiar splitting function $P_{gq}(z)$, which is 
precisely the driving term of the quark-jet fragmentation function.  
Consequently, the contribution (\ref{eq:8.1.2}) must be treated 
as a fragmentation of
the scattered quark into the quark and gluon, $q' \to qg$.  
The quark pole contribution will dominate if 
\beq
\bk_{\perp}^2 \ll (\bp-\bDelta)^2 =\bp_q^2.
\label{eq:8.1.3}
\eeq
From the experimental point of view, 
the corresponding final state is a monojet of the 
transverse momentum $\bDelta$. The transverse momentum of 
such a monojet will be compensated
by an away jet produced at midrapidity or the nucleus
hemisphere of $pA$ collisions.

In terms of  Feynman diagrams of Fig. 
\ref{fig:QuarkGluonDijetExcitation} - for the free-nucleon target
one takes the single-gluon exchange, - 
the monojet production is a property of the diagram (c). 
Indeed, the cross section 
(\ref{eq:8.1.2}) is proportional to precisely the differential cross
section of quasielastic scattering of the projectile quark
off the nucleon target - the latter is evidently proportional to 
the unintegrated gluon density of the target proton  $f(\bDelta)$.
The two classes of Feynman diagrams in Fig. \ref{fig:QuarkGluonDijetExcitation}, 
(b) and (c), are integral parts of the gauge-invariant description of the
QCD Bremsstrahlung excitation of the $qg$ state. Still, the isolation 
of the pole contribution from the gauge-invariant combinations 
$$
\Bigl|\Psi(1;z,\bp-\bDelta) -\Psi(z,\bp -z\bDelta) \Bigr|^2
=\Bigl|\Psi(1;z,\bp_q) -\Psi(z,\bp -z\bDelta) \Bigr|^2
$$ in
(\ref{eq:5.8}), and of the monojet contribution to the generic 
dijet cross section  wouldn't conflict gauge invariance. In order to
conform to the jet-finding algorithms, the production of the quark and gluon 
within the jet-defining cone must be treated as a fragmentation of the 
monojet; if the
azimuthal angle between the quark and gluon is larger than
the jet-defining angle, the two jets must be viewed as 
independent ones. The combination of the wave functions, which enters 
the excitation
of the sextet and 15-plet final states, see Eq.~(\ref{eq:7.3.1}),
has the form $$|\Psi(z,\bp)-\Psi(z,\bp - \bDelta)|^2 \propto
{(\bp -\bp_q)^2 \over \bp^2\bp_q^2}$$ and is finite for all orientations
of the quark and gluon jets.

The quark-tagged pQCD gluon Bremsstrahlung considered here is already the higher
order process, the lowest order pQCD process in $qN$ interaction 
is the radiationless quasielastic scattering of the quark. Naive application of
fragmentation $q'\to qg$ to this lowest order process would evidently
lead to a double counting, because the fragmentation is manifestly
a monojet part of our dijet cross section. The integration over the 
gluon momentum $\bk_{\perp}$ in the inclusive cross section would 
yield the  familiar collinear logarithm, which must be 
reabsorbed into the definition
of the fragmentation function at the starting scale.
Simultaneously, one must include the
virtual radiative correction to the radiationless quasielastic scattering of the 
incident quark off the target nucleon. 
The treatment of these virtual corrections to quasielastic scattering
and elimination of double counting 
go beyond the scope of the present study and will be addressed elsewhere.
We only want to comment that if one would insist on the description of
monojets in terms of the fragmentation of the quark, then
the interplay of the virtual correction
to the radiationless quasielastic scattering and of the collinear 
logarithm in the monojet component of the the dijet cross section may 
entail a departure of the fragmentation function from that defined
in the $e^+e-$ annihilation.

%%%%%%%%%%%%  section 8.2

\subsection{Monojets from dijets off a nuclear target}

The presence of the monojet pole (\ref{eq:8.1.1}) in the nuclear
dijet cross section (\ref{eq:6.5.1}) is manifest:
\bea
&&\left. {d\sigma_{A}(q^*\to qg) \over d^2\bb dz d\bp d{\bDelta}}\right|_{monojet} = 
\frac{1}{
2(2\pi)^2} T(\bb) \phi(\bb,x,\bDelta) \Bigl|\Psi(z,\bp -z\bDelta)
\Bigr|^2\, .
\label{eq:8.2.1}
%\vskip{-0.5cm}
\eea 
It factorizes precisely as the free-nucleon cross section: the
differential cross section of quasielastic quark-nucleus scattering,  
proportional to the unintegrated collective gluon density of the nucleus,
times
the fragmentation of the scattered quark to the gluon and quark 
given by $|\Psi(z,\bp -z\bDelta)|^2$, which does not depend on
the target. 
However, 
the virtual radiative correction to the radiationless
quasielastic scattering
of the incident quark off the target nucleus and the elimination of 
double counting are likely to depend on the acoplanarity 
momentum $\bDelta$ and the shape of the 
collective nuclear glue $\phi(\bb,x,\bDelta)$. Should this be the case,
such a dependence could be reinterpreted as a nuclear modification of the 
fragmentation function; this issue will be addressed 
elsewhere.

As it was the case for the free-nucleon target, excitation of the 
sextet and 15-plet final states is free of the monojet singularities.
To be more precise, the wave-function singularities in the integrand of the 
sextet and 15-plet contribution to (\ref{eq:6.5.1}) occur in the
intermediate state, at $\bp-\bkappa_1-\bkappa=0$ and $\bp-\bkappa_1=0$.
However, they are integrated out in the observed dijet
cross section. It is still instructive to look at the effect
of these singularities in 
the monojet kinematics
$\bDelta^2\gg \bp^2 \gsim Q_A^2(\bb)$. 

Consider first the contribution from the intermediate pole 
at $\bp-\bkappa_1=0$. The relevant $\bkappa_i$ integrations are of the form
\bea
&&
\int d^2\bkappa_1 d^2\bkappa f(x,\bkappa)
\Phi(1-\beta,\bb,x,\bDelta -\bkappa_1 -\bkappa)
\Phi(\beta+C_2(1-\beta),\bb,x,\bkappa_1)\nonumber\\
&\times&|
\Psi(\beta;z,\bp -\bkappa_1)|^2 \nonumber\\
&=& \Phi(\beta+C_2(1-\beta);\bb,x,\bp)\int^{\bp^2} d^2\bk|\Psi(\beta;z,\bk)|^2 
\nonumber\\
&\times&\int d^2\bkappa f(x,\bkappa)\Phi(1-\beta,\bb,x,\bDelta -\bp -\bkappa)
\label{eq:8.2.2}
\eea 
For the considered hard jets 
\beq
\Phi(1-\beta,\bb,x,\bDelta -\bp -\bkappa) = {1\over 2}(1-\beta)T(\bb)f(\bDelta -\bp -\bkappa)
\label{eq:8.2.3}
\eeq
and the convolution in (\ref{eq:8.2.2}) equals \cite{NSSdijet,Nonlinear}
\bea
\int d^2\bkappa f(x,\bkappa)\Phi(1-\beta,\bb,x,\bDelta -\bp -\bkappa)=(1-\beta)T(\bb)f(\bDelta -\bp)\,.
\label{eq:8.2.4}
\eea 
The resulting contribution from the intermediate pole of the wave function 
at $\bp-\bkappa_1=0$ is proportional to
\bea
T^2(\bb) f(\bDelta -\bp)f(\bp) P_{gq}(z) = T^2(\bb) f(\bp_g)f(\bp_q) P_{gq}(z)
\label{eq:8.2.5}
\eea  
and has the form of the product of the differential cross sections
of independent quasielastic scattering of the quark and gluon
fragments of the incident quark.
 It does
not depend on the azimuthal angle between the quark and gluon jets at all,
and has no collinear singularity.
A similar situation has been found to occur in our previous study of the
production of hard quark-antiquark dijets
in $\pi A$ collisions \cite{PionDijet}.
The contribution from the pole at $\bp-\bkappa_1-\bkappa=0$ is entirely similar.

\section{Conclusions}

We presented a derivation of nuclear modifications of the  
quark-gluon production in quark-nucleus collisions. Our
principal result is the nonlinear $k_{\perp}$-factorization
relation (\ref{eq:6.5.1}).  The derived dijet cross
section can be decomposed into three major contributions.
The excitation of $qg$ dijets in higher - sextet and 15-plet -
color
representations gives rise to the sixth order nonlinearity in 
gluon fields, compared to the fifth order nonlinearity
for $q\bar{q}$ dijets in DIS.
A part of the nonlinearity comes from the free-nucleon gluon density
which emerges in all instances of excitation of higher color representations
(see also the related discussion of the $1/(N_c^2-1)$ expansion in 
Ref. \cite{Nonlinear}). The matrix elements of transitions from lower to higher 
color representations are suppressed at large $N_c$, but this
suppression is compensated for by the 
large number of states in higher
representations. The coherent diffraction, in which the final 
dipole is produced in exactly the
same color state as the incident quark, is not suppressed by large $N_c$,
but because of the color-nonsinglet incident partons the diffractive
contribution is suppressed by an overall nuclear attenuation and
will only come from collisions at the diffuse edge of a nucleus.
A new feature
of $qA$ collisions in contrast to DIS is inelastic production of $qg$ states
in the same color representation as the incident parton. Such 
color rotations within the same representation are not
suppressed at large $N_c$. This contribution has the form which superficially
looks like satisfying the linear $k_{\perp}$-factorization in 
terms of the collective nuclear gluon density. However, it contains
the nuclear-distorted wave function of the $qg$ Fock state
and, consequently, is a cubic
functional of the collective nuclear glue.

The above three components of the dijet cross section differ by more
than the degree of the nonlinearity. The coherent diffractive mechanism
and the excitation of quark-gluon dijets in the same color
representation as the incident quark are 
explicitly calculable in terms of the collective nuclear glue 
of Eq. ~(\ref{eq:6.2.1}) which is defined for the whole 
nucleus. This is not the case for the excitation of quark-gluon
dijets in higher color multiplets. It is proportional to the
unintegrated gluon density in the free nucleon. The 
coherent initial 
state interaction, before the excitation of higher color multiplets
at the depth $\beta$ of the nucleus, must be described in terms 
of the unintegrated collective glue (\ref{eq:6.2.6}) defined
for the slice $\beta$ of the nucleus. 
Coherent distortions of the $qg$ wave function 
are complemented by incoherent broadening of the incident 
quark transverse momentum distribution
in the same slice of the nucleus. 
Likewise, the final
state interactions after the excitation of higher multiplets
must be described in terms  of the unintegrated collective 
glue defined for the slice $(1-\beta)$ of the nucleus. 
This reinforces the point \cite{Nonlinear} that 
hard processes in a nuclear environment can not be
described in terms of a nuclear gluon density defined for
the whole nucleus, as it was advocated, for instance, within
the Color Glass Condensate approach \cite{CGC}. 
Furthermore, besides the collective nuclear glue defined for
color-singlet quark-antiquark dipole, there emerges a new
nuclear gluon density which depends on the Casimir operators 
of higher quark-gluon color representations, i.e., gluon field 
of the nucleus must be described by a density matrix in the 
space of color representations. Based on a  comparison of the 
excitation of quark-gluon dijets in quark-nucleon collisions to
the excitation of quark-antiquark dijets in DIS and gluon-nucleus
collisions, we formulated four universality classes for
nonlinear $k_{\perp}$-factorization.

The representation for the dijet cross section similar to our
master formula (\ref{eq:2.12}) has been discussed recently by
several authors \cite{Blaizot,Kovchegov,Raju}, but these works stopped
short of the solution of the coupled-channel intranuclear evolution 
for the for 4-parton state. Although major ingredients for the
diagonalization of the four-body $\textsf{S}$-matrix are found in our
earlier work on dijets in DIS \cite{Nonlinear}, the case of
the $qg$ dijets has its own tricky points. For this reason, we
felt it imperative to present full technical details of this
diagonalization. 

The emphasis of the present communication was 
on the formalism, the numerical applications will be reported
elsewhere. The nuclear coherency 
condition, $x \lsim x_A \approx 0.1\cdot A^{-1/3}$, 
restricts the applicability 
domain of our formalism to the forward part of the 
proton hemisphere of $pA$ collisions at 
RHIC. Although the required coherency condition does not hold for the 
mid-rapidity dijets studied so far at RHIC \cite{STAR}, our
predictions could be tested after the detectors at RHIC II 
will be upgraded to cover the proton fragmentation region \cite{RHIC_II}.

This work was partly supported by the grant DFG 436 RUS 17/101/04.

\end{document}